\documentclass[letterpaper, twocolumn,superscriptaddress, showpacs,preprintnumbers,amsmath,amssymb] {revtex4}
\usepackage{graphicx}
\usepackage{dcolumn}
\usepackage{bm}
\usepackage{hyperref}
\hypersetup{
    colorlinks,
    citecolor=red,
    filecolor=blue,
    linkcolor=blue,
    urlcolor=blue
}

\begin{document}

\title{Heat and spin transport in a cold atomic Fermi gas}

\author{Hyungwon Kim}

\affiliation{Physics Department, Princeton University, Princeton, NJ 08544, USA}

\author{David A. Huse}
\affiliation{Physics Department, Princeton University, Princeton, NJ 08544, USA}

\begin{abstract}
Motivated by recent experiments measuring the spin transport in ultracold
unitary atomic Fermi gases [Sommer et al. Nature (London) 472 201 (2011); Sommer et al. New J. Phys. 13 055009 (2011)], we explore the theory of spin and heat
transport in a three-dimensional spin-polarized atomic Fermi gas.
We develop
estimates of spin and thermal diffusivities
and discuss magnetocaloric effects, namely the the spin Seebeck and spin Peltier effects.
We estimate these transport coefficients using a Boltzmann kinetic equation in the classical regime and
present experimentally accessible signatures of the spin Seebeck effect.
We study an exactly solvable model that
illustrates the role of momentum-dependent scattering in the
magnetocaloric effects.
\end{abstract}


\maketitle

\section{Introduction}

The transport properties of condensed matter systems are often measured by driving currents
externally and measuring the resulting voltages or temperature differences.  In cold atomic
gas clouds, on the other hand, transport is more often measured by setting up
transient out-of-equilibrium initial conditions
and measuring the subsequent relaxation towards equilibrium \cite{sommer1,sommer2,demarco,strohmaier,trenkwalder,hulet}.
In the approximation that the cloud is isolated and has an infinite lifetime (no loss of atoms or exchange
of energy with any degrees of freedom outside of the gas cloud), the conserved currents of interest
include the energy current and currents of each of the atomic species present.  In the absence of
optical lattices or random potentials that violate momentum conservation, one can also ask about
the transport of momentum (viscosity).

In this paper we consider the diffusive transport of heat and of atoms.  We mostly focus on
the case of a two-species Fermi gas with only inter-species contact interactions, but start with a somewhat more
general discussion here.  The system may, in addition to diffusive transport, also have underdamped
or propagating sound or other ``collective'' modes.  A gas cloud in a smooth trap will have such
sound modes, with the longest-wavelength sound modes being the often-discussed collective modes of
the cloud's oscillations
within the trap.  Here we consider a gas cloud in a smooth trap, with the cloud at global {\it mechanical} equilibrium, so that
any pressure gradients in the cloud are sufficiently balanced by trapping forces that no underdamped sound or
collective modes are excited.  We also assume that the cloud is
everywhere near {\it local} thermodynamic equilibrium, so the
local temperature $T({\bf r})$ and local chemical potentials $\mu_i({\bf r})$ can be defined.
However, the cloud may still
have gradients in the local temperature and in
the local
chemical potentials of the various species of atoms.
If the equilibrium equation of state of the system is known
(for the unitary Fermi gas, see \cite{ku, nascimbene}) then
measurements of the local densities of each species allows these gradients of $T$ and the $\mu_i$'s
to be measured.  Thus, for example, the local densities can be used as local thermometers to allow
a measurement of the thermal diffusivity by an approach similar to that used in \cite{sommer1,sommer2}
to measure the spin diffusivity (but with an initial temperature gradient instead of a composition gradient).

The transport currents that we
examine in this paper are those that arise in linear response to these gradients.
In a trap, convection currents may also appear in linear response, as temperature and/or
composition gradients may produce density inhomogeneities, and the ``heavier'' regions of
the gas cloud will sink towards the bottom of the trap while the ``lighter'' regions rise.  These
convection currents are damped by the viscosity.  Convection will be strongest in wide clouds and
should be much weaker in high-aspect-ratio clouds with the gradients in $T$ and the $\mu_i$'s
oriented along the long axis of the cloud.
In most of this paper, for simplicity we consider a gas in
a spatially uniform potential, so such convection currents do not appear in linear response.
Then mechanical equilibrium is indeed a sufficient condition to have a convectionless gas \cite{smith}.

Quite generally, a temperature gradient drives a heat current and a gradient of chemical potential difference
drives a composition (``spin'') current, consisting of opposing currents of the two (or more) atomic species.
We call these ``direct'' responses to a temperature gradient and a gradient of chemical potential difference ``primary currents''.
In addition to these ``primary currents'', there are the magnetocaloric currents, namely spin Seebeck currents
(spin currents induced by a temperature gradient) and spin Peltier currents (heat currents induced by gradients of chemical potential difference).
These magnetocaloric effects have been one of the central research topics in the field of spintronics \cite{wolf}.
The spin Seebeck effect \cite{kimura,uchida} and the spin Peltier effect \cite{flipse} have already been observed in condensed matter systems,
while they are yet to be detected in cold atomic clouds.
In this paper we discuss the origin and the physics of these effects in a cold atomic Fermi gas,
and estimate how large these effects can be in realistic experiments.

Note that Ref. \cite{wong} discusses a rather different
situation that they are also calling the ``spin Seebeck effect'': they consider an unpolarized gas with the two species
at different temperatures (thus {\it not} in {\it local} equilibrium) and a gradient in this temperature difference.
Also, Ref. \cite{grenier} studies a different system, a ``two terminal geometry'',
considering transport through a narrow constriction between two reservoirs held at different temperatures and chemical potentials.
For this constriction, they discuss ``off-diagonal'' elements in the transport matrix which they call effective Seebeck and Peltier effects.

This paper begins with a general discussion of the two-species universal Fermi gas and an introduction of transport coefficients,
spin and thermal diffusivities, spin Seebeck effect, and spin Peltier effect.
Then we present rough estimates of diffusivities and determine the signs of the spin Seebeck and spin Peltier effects based on physical arguments.
In the following section, these qualitative descriptions are justified by approximate solutions of the Boltzmann transport equation in the classical regime.
We then compute experimentally verifiable signals of the spin Seebeck effect.
These results are tested against an exactly solvable model, namely atoms with a ``Maxwellian'' scattering cross section.
We also use the Kubo formula to derive various general relations among the
transport currents and coefficients.

The various effects discussed in this paper are probably most accessible
experimentally for the unitary Fermi gas at temperatures of order the Fermi temperature, where the diffusivities
are at their smallest, so the diffusive relaxation towards equilibrium is slowest and most easily studied.

We set Boltzmann's constant $k_B = 1$ but explicitly keep Planck's constant $\hbar$.

\section{Universal Fermi Gas}

The universal Fermi gas is a two-species Fermi gas with only contact ($s$-wave) inter-species interactions
that is realized to a very good approximation in recent experiments with ultracold atoms \cite{kz,navon}.
We consider such a gas in three-dimensional space, with the two species having equal mass $m$.  There is
no optical lattice, only a possible smooth trap potential.
The interaction is specified by the scattering length $a$, which can be set to any
value in experiments by tuning through a Feshbach resonance \cite{regal,ohara}.  For $a=0$ this is the
standard textbook noninteracting Fermi gas, while for weakly attractive $a$ it is
very close to the
model used by Bardeen, Cooper and Schrieffer (BCS) to explain superconductivity.
Indeed, this system shows paired-fermion superfluidity at low temperatures.
The limit of infinite $|a|$ is the strongly-interacting unitary Fermi gas.

As is standard, we call the majority species with number density $n_{\uparrow}$ ``up'',
and the minority species ``down'', $n_{\downarrow}\leq n_{\uparrow}$.
The total number density $n=n_{\uparrow}+n_{\downarrow}$, together with $m$
and $\hbar$ set the characteristic length, time and energy scales.
The scaled dimensionless properties of this universal Fermi gas then depend
on only three dimensionless parameters, which can be chosen to be $1/(k_Fa)$, $T/T_F$ and
the polarization $p=(n_{\uparrow}-n_{\downarrow})/n$.
We use a convention that the Fermi wavenumber and temperature $k_F$ and $T_F$ are defined by the total density,
so that at high polarization $T_{F\downarrow}\ll T_F \approx T_{F\uparrow}$
and $k_{F\downarrow}\ll k_F \approx k_{F\uparrow}$.  This universal Fermi gas
has a variety of regimes of behavior
: The polarization $p$ can be low or zero
so $n_{\downarrow}\cong n_{\uparrow}$ or it can be near one so $n_{\downarrow}\ll n_{\uparrow}$.
The temperature can be higher, $T>T_{F\uparrow}$, or lower, $T<T_{F\downarrow}$, than both
Fermi temperatures or, for $p>0$ it can be in between them, $T_{F\downarrow}<T<T_{F\uparrow}$.
The scattering can be near unitarity so $|k_Fa|$ is of order one or more, or it can be far
from unitarity so $|k_Fa|\ll 1$.  At high $T$ it also matters whether $|a|$ is larger or smaller
than the thermal de Broglie wavelength $\lambda\equiv \sqrt{2\pi\hbar^2/m T} \sim T^{-1/2}$.  
There are
also important differences between the $a<0$ (BCS side of the Feshbach resonance) and
$a>0$ (BEC side) regimes.

\section{Transport}

The conservation laws of this Fermi gas are: total energy ($E$), total momentum (${\bf \Pi}$),
and the total number of each of the species ($N_\uparrow$ and $N_\downarrow$).
The viscosity measures the transport of momentum, which we mostly do not consider here.  Thus we consider primarily
the transport of atoms and of heat.  If there is a nonuniform pressure in the system that is not balanced by
a trapping potential, the gas will accelerate
and this will produce free motion or propagating sound waves.  Here we
consider the diffusive spin and heat transport in a gas with no trapping potential and
spatially uniform pressure $P$, so it is at mechanical equilibrium.  The gas is near {\it local}
thermodynamic equilibrium, but with possible weak gradients in
the local temperature and/or the spin polarization.  A smooth trap potential may be added via
the local density approximation (LDA).

In general, an inhomogeneity of the Fermi gas consists of gradients in the local temperature and
of the local densities of the two atomic species.
Mechanical equilibrium imposes a constraint on these gradients and thus there are only two
independent linear combinations of the three gradients. One way of
describing the diffusive dynamics is in terms of the atomic densities $n_{i}$ and the currents
${\bf j}_i$ of each species $i=\uparrow$,$\downarrow$, leaving the
temperature and the heat current implicit, since they are dictated by the equilibrium equation of state,
e.g., $T(P,n_{\uparrow},n_{\downarrow})$.  This description of the transport has the virtue that it is in
terms of what appears to be the most accessible local observables in experiment, namely the local densities of each species.
Since the system in the absence of a trapping potential
is Galilean-invariant, we have a certain amount of flexibility in what inertial frame we use to specify
the currents.
For most of this work, we consider the frame where the center of mass of the whole cloud is at rest
and let the gas have long-wavelength temperature, density and/or composition modulations, but always with a
spatially uniform pressure.  The diffusive currents
are related to the density gradients as
\begin{equation}\label{diffusion}
\begin{pmatrix}
\mathbf{j}_{\uparrow}\\
\mathbf{j}_{\downarrow}
\end{pmatrix}
=-
\begin{pmatrix}
{\frak D}_{\uparrow\uparrow} && {\frak D}_{\uparrow\downarrow}\\
{\frak D}_{\downarrow\uparrow} && {\frak D}_{\downarrow\downarrow}
\end{pmatrix}
\begin{pmatrix}
\nabla n_{\uparrow}\\
\nabla n_{\downarrow}
\end{pmatrix}~.
\end{equation}

The diffusion matrix in (1) has two eigenmodes.
At zero polarization, the symmetry between
$\uparrow$ and $\downarrow$ implies ${\frak D}_{\uparrow\uparrow} = {\frak D}_{\downarrow\downarrow}$
and ${\frak D}_{\uparrow\downarrow} = {\frak D}_{\downarrow\uparrow}$.
Therefore one eigenmode is odd under exchanging species,
$\frac{1}{\sqrt{2}}(1,-1)$; the current in this odd mode carries only spin and no
net density or energy. The other eigenmode is the even mode, $\frac{1}{\sqrt{2}}(1,1)$;
the current in this even mode carries both net density and energy, but no spin.
Away from zero polarization, when $n_{\uparrow}\neq n_{\downarrow}$,
we no longer have this symmetry between species.
The diffusive eigenmodes are then no longer purely spin or purely not spin, instead they are mixtures, thus
producing the spin Seebeck and Peltier effects.  The eigenmode where the currents of the
two species are in opposite directions we will call the ``spin'' mode with
diffusivity ${\frak D}_s$, while the other
mode where they are parallel we will call the ``thermal'' (or heat) mode with diffusivity ${\frak D}_T$.

Another standard representation of the transport matrix in terms of the heat current ${\bf j}_{heat}$ and spin current ${\bf j}_{spin}$
is the following:
\begin{equation}\label{diffusion_Tmu}
\begin{pmatrix}
\mathbf{j}_{heat} \\
\mathbf{j}_{spin}
\end{pmatrix}
=-
\begin{pmatrix}
\kappa && P_s \\
S_s && \sigma_s
\end{pmatrix}
\begin{pmatrix}
\nabla T \\
\nabla (\mu_\uparrow - \mu_\downarrow)
\end{pmatrix}
~,
\end{equation}
where $\kappa$ is the thermal conductivity and $\sigma_s$ is the spin conductivity.
$S_s$ and $P_s$ are the spin Seebeck and Peltier coefficients, respectively, and they are related by the Onsager relation, $P_s = T S_s$.
This matrix explicitly shows the direct responses (diagonal elements) and magnetocaloric effects (off-diagonal elements),
and is in the form that is given by the Kubo formula, as we discuss below.

The currents in (2) must be defined properly so that they are the
transport currents, namely the currents of heat and spin relative to the average local motion of the gas.
Let the local current density of atoms be ${\bf j}_n = {\bf j}_\uparrow + {\bf j}_\downarrow$.  These atoms carry
the average heat, $s T$, where $s$ is the average entropy per particle, and the average spin polarization $p$.
Therefore the local heat and spin transport currents are
\begin{align}
{\bf j}_{heat} &= {\bf j}_{\epsilon}  - \mu_\uparrow {\bf j}_\uparrow - \mu_\downarrow {\bf j}_{\downarrow} - sT {\bf j}_n\label{heat}\\
{\bf j}_{spin} &= \frac{1}{2}\left({\bf j}_\uparrow - {\bf j}_\downarrow - \frac{(n_\uparrow - n_\downarrow)}{n}{\bf j}_n\right)\nonumber\\
&= \frac{n_\downarrow}{n}{\bf j}_\uparrow - \frac{n_\uparrow}{n}{\bf j}_\downarrow ~,\label{spin}
\end{align}
where ${\bf j}_{\epsilon}$ is the local energy current.
Since the above currents measure only the transport relative to the average motion of the gas, they are {\it reference frame independent}.
For more details of definitions of currents, see e.g. Ref \cite{chaikin}.

If the full equation of state of the system is known, then measurements of the pressure and the local densities can be converted
to local temperatures and chemical potentials.  But the local density $n$ and polarization $p$ are directly observable without requiring
knowledge of the equation of state, so yet another convenient form of the transport equations is

\begin{equation}\label{diffusion_Tp}
\begin{pmatrix}
\mathbf{j}_{heat} \\
\mathbf{j}_{spin}
\end{pmatrix}
=-
\begin{pmatrix}
\kappa' && P_s' \\
S_s' && D_s
\end{pmatrix}
\begin{pmatrix}
\nabla T \\
n \nabla p
\end{pmatrix}
~.
\end{equation}
At $p=0$, we have $\kappa'=\kappa$ and $D_s={\frak D}_s$, but when $p \neq 0$ these quantities in general differ due to the mixing between
spin and heat transport.  It is possible that $S_s'$ is the most directly accessible version of the spin Seebeck coefficient: if one
can set up an initial condition at mechanical equilibrium and local thermodynamic equilibrium with a temperature gradient but no polarization gradient
and then measure the resulting spin current, this is a measurement of $S_s'$ and does not require knowledge of the equation of state.

Note that the three representations, Eq. (\ref{diffusion}), Eq. (\ref{diffusion_Tmu}) and Eq. (\ref{diffusion_Tp}) are related by the equation of state, the mechanical equilibrium condition, and definitions of spin and heat currents. Hence, they are {\it equivalent}.

\section{Diffusivities}

Let's first present rough
``power-counting'' estimates of the spin and thermal diffusivities, $D_s$ and $D_T$, respectively.
At the level of power-counting the differences between the various possible definitions of these diffusivities
are small and are ignored here.
Previous work \cite{sommer1,bruun1} on the unpolarized gas ($p=0$) shows that $D_s$ for $T>T_F$
is the larger of $\frac{\hbar}{m}(\frac{T}{T_F})^{3/2}$
and $\frac{\hbar}{mk_F^2a^2}\sqrt{\frac{T}{T_F}}$.
In the recent experiment, which was performed at
unitarity \cite{sommer1}, this high-$T$ behavior is observed, with significant deviations
apparently beginning between $T=2T_F$ and $T_F$ as $T_F$ is approached from above.
Staying in this high-$T$ regime, as we move to high polarization ($p$ near 1) at a given $n$ and $T$,
the scattering time of species $i$ is roughly $\tau_i \sim \frac{1}{n_{j}\sigma v_r}$ where $i \neq j$
and $v_r = |{\bf v}_\uparrow - {\bf v}_\downarrow|$ and $\sigma$ is the $s-$wave scattering cross section (see Eq. (\ref{general_cross_section}))
evaluated at a typical value of momentum.
Thus, the scattering time of the down atoms $\tau_{\downarrow}$ decreases by only a factor of two due to
the increase of the density $n_{\uparrow}$ of the up atoms that
they scatter from.  The up atom scattering time $\tau_{\uparrow}$, on the other hand, increases by a
factor of $n_{\uparrow}/n_{\downarrow}$ as the down atoms that they scatter from become dilute.
At high polarization, the spin current consists of the down atoms moving with
respect to the up atoms at typical speed $v_{\downarrow}\sim\sqrt{T/m}$,
so $D_s\sim v^2_{\downarrow}\tau_{\downarrow}$ is not strongly
polarization dependent for $T>T_{F\downarrow}$; the experimental results \cite{sommer1,sommer2} are
consistent with this.  The heat, on the other hand, is mostly carried by the up atoms at
high polarization, resulting in $D_T \sim D_s n_{\uparrow}/n_{\downarrow}$, a relation between the two diffusivities
that appears to remain true at high polarization for all $T$ away from the superfluid phases.  At $p=0$ and high $T$ the two
diffusivities are comparable, but $D_T$ remains larger than $D_s$ because a single $s$-wave scattering
event completely randomizes the total spin current carried by the two atoms,
while the component of the heat current carried by their center of mass is preserved.
Thus
it appears that the heat mode always diffuses faster than the spin mode.

Moving towards lower $T$, let's next pause at $T=T_{F\uparrow}$, noting that
here $v_{\uparrow}\sim v_{\downarrow}\sim\sqrt{T/m}$, and $\tau_{\downarrow}$ is the larger
of $\frac{\hbar}{T}$ and $\frac{\hbar}{Tk_F^2a^2}$.  For the polarized gas
$\tau_{\uparrow}\sim\tau_{\downarrow}n_{\uparrow}/n_{\downarrow}$.

We next (still just power-counting) look at the polarized gas
in the intermediate temperature regime $T_{F\downarrow}<T<T_{F\uparrow}$ where the majority
atoms are degenerate ($v_{\uparrow}\sim\sqrt{T_{F\uparrow}/m}$),
while the minority atoms are not ($v_{\downarrow}\sim\sqrt{T/m}$).
Some changes from the high-T regime are: only up atoms with energy within
$\sim T$ of $T_{F\uparrow}$ are involved in the scattering and
all but a fraction $T/T_{F\uparrow}$ of the final states of the
scattering are Pauli-blocked due to the degeneracy of the up atoms.
This increases $\tau_{\downarrow}$ by a factor of $(T_{F\uparrow}/T)^2$,
and $\tau_{\uparrow}$ by a factor of $T_{F\uparrow}/T$, compared to their
values at $T=T_{F\uparrow}$.
Thus, $\tau_{\downarrow}$ is the larger of $\frac{\hbar}{T_{F\uparrow}}\left(\frac{T_{F\uparrow}}{T}\right)^{2}$ and $\frac{\hbar}{T_{F\uparrow}k^{2}_{F\uparrow}a^{2}}\left(\frac{T_{F\uparrow}}{T}\right)^{2}$.
As a result, $D_s \sim v_{\downarrow}^{2}\tau_{\downarrow}$ is the larger of
$\frac{\hbar T_{F\uparrow}}{mT}$ and $\frac{\hbar T_{F\uparrow}}{mTk_{F\uparrow}^2a^2}$, while
$D_T$ is again larger than $D_s$ by a factor $n_{\uparrow}/n_{\downarrow}$.
This estimate of $D_s$ is consistent with a previous quantitative calculation \cite{bruun2,hyungwon}.
Note that the temperature dependence of $D_s$ crosses over from a decreasing function
of $T$ at low $T$ to an increasing function at high $T$.  The recent measurements \cite{sommer2}
of the spin drag in a polarized unitary gas show this crossover occurring at
roughly $T=0.4 T_{F\uparrow}$.
On the BEC side of the Feshbach resonance, the minority atoms bind in to bosonic
Feshbach molecules at low enough $T$.  But as long as these molecules remain
nondegenerate and thus not superfluid, the above estimates of the diffusivities should hold.

For $T<T_{F\downarrow}$, the minority atoms become degenerate.  This leads to
superfluidity on the BEC side of the Feshbach resonance as well as at low polarization
near unitarity.  But there are regimes on the BCS side of the resonance as well as
near unitarity at high polarization where the minority atoms
(near unitarity strongly ``dressed'' as polarons) form a degenerate Fermi
gas.  Here the important change from the intermediate temperature regime
at the level of ``power-counting'' is that only
minority atoms with energy within $\sim T$ of $T_{F\downarrow}$ are involved in the
scattering and they have momentum $k_{F\downarrow}$ instead of a thermal momentum.
This increases the diffusivities by a factor of $T_{F\downarrow}/T$, so the spin
diffusivity in these degenerate Fermi liquid regimes is the larger of
$\frac{\hbar T_{F\uparrow}T_{F\downarrow}}{mT^2}$ and
$\frac{\hbar T_{F\uparrow}T_{F\downarrow}}{mT^2k_{F\uparrow}^2a^2}$.
We expect that $D_T$ is still greater than $D_s$ by a factor of $n_{\uparrow}/n_{\downarrow}$
but this question should be examined more carefully within Fermi liquid theory.

At low temperature, the
polarized Fermi liquid may become a $p$-wave superfluid, with pairing within one species
mediated by the attraction to the other species \cite{bulgac1}. Or it may have a Fulde-Ferrell-Larkin-Ovchinnikov (
FFLO) phase with Cooper pairs of nonzero total momentum \cite{sheehy, bulgac2, liao}.
In the superfluid phases, the thermal diffusivity should diverge to infinity,
as heat is carried ballistically by second sound modes.  The spin diffusivity
presumably remains finite in the superfluid phases,
although, as we discuss
below, the spin Seebeck effect appears to
generally be divergent in a polarized superfluid.

\section{Spin Seebeck and Spin Peltier Effects}

More challenging to estimate than these spin and
thermal diffusivities are the effects that mix spin and heat transport,
namely the spin Seebeck and the spin Peltier effects.
Here we present ``simple'' arguments for the signs of these effects in two regimes:
(1) far away from unitarity for all temperature ranges, and (2) at unitarity in the classical regime.
Here we always consider a spin-polarized gas, since these ``magnetocaloric'' effects vanish by symmetry in the case of an
unpolarized gas where the two species also have equal mass.

(1) Well away from unitarity ($|k_F a| \ll min\{1, \sqrt{T_F/T}\}$),
the scattering cross section is essentially $a^2$ and independent of momentum for all temperatures,
since the interaction is weak and the atoms are not thermally excited to $\lambda < |a|$.
Generally, the scattering rate is proportional to the {\it cross section} $\times$ {\it relative speed}.
Thus, in this (low energy) regime, the scattering rate is $\sim a^2 |{\bf v}_\uparrow - {\bf v}_\downarrow|$.
This implies minority atoms will scatter more frequently with the majority atoms of higher energy than majority atoms of lower energy
since higher-energy majority atoms have higher relative speed.
Consequently, the direction of the minority current is along the flow of higher-energy (``hot'') majority atoms
and thus parallel to the heat current.
For a uniformly spin-polarized gas with a temperature gradient, where the spin Seebeck effect occurs,
the primary current is the heat current transporting ``hot'' majority atoms from the hot region to the cold region
and transporting ``cold'' majority atoms from the cold region to the hot region.
The fact that the minority current is aligned with that of the ``hot'' majority atoms means
the direction of the net spin current (spin Seebeck current) is opposite from that of the heat current.
Thus, the initially cold region becomes less polarized due to minority atoms transported by the spin Seebeck current.

\begin{figure}
\includegraphics[width=3.50in]{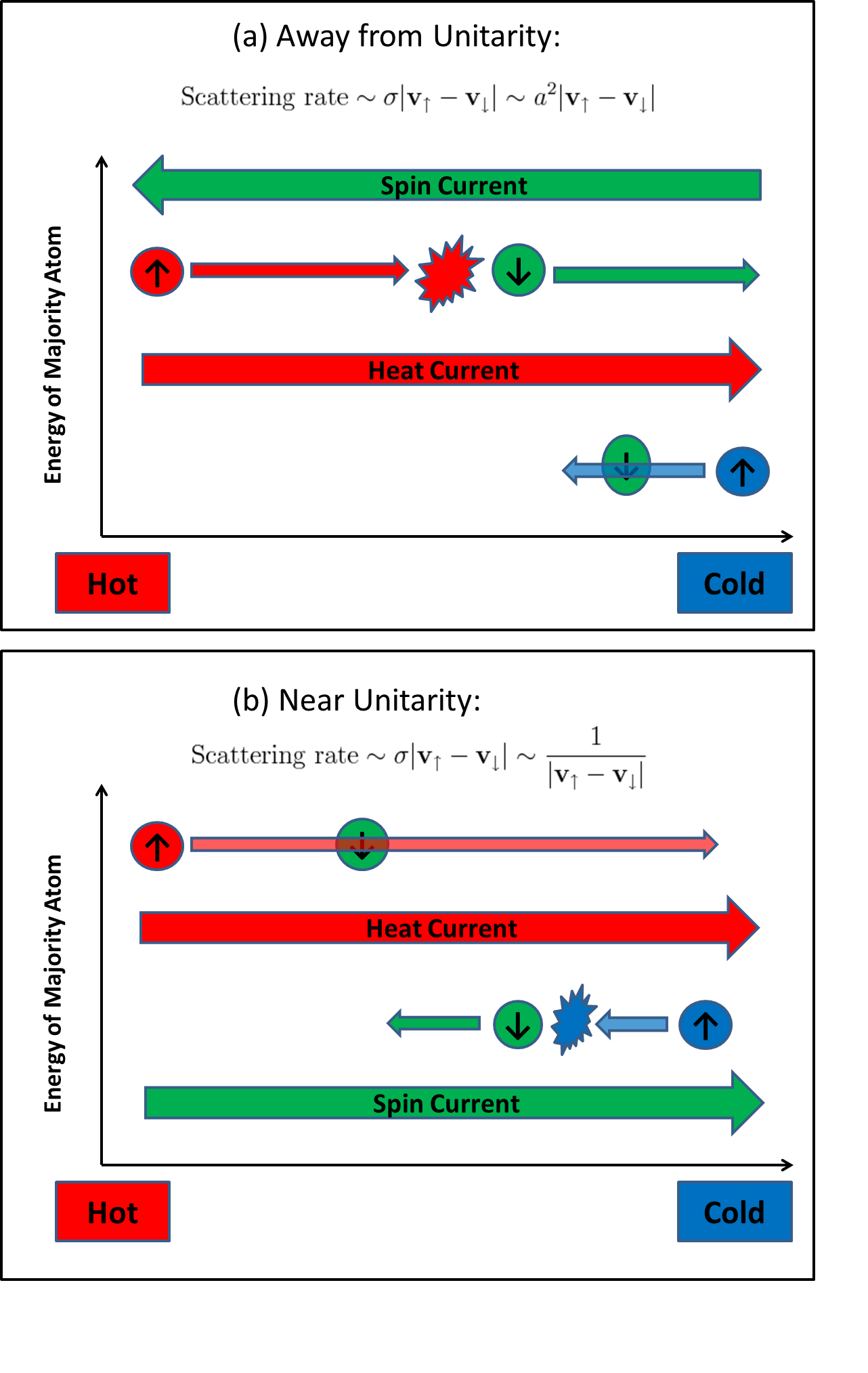}
\centering
\caption{(Color online) Illustration of the spin Seebeck effect. (a) Far away from unitarity, the scattering rate is proportional to the relative speed. Thus, ``hot'' majority ($\uparrow$) atoms (top) collide more often with minority ($\downarrow$) atoms than do ``cold'' majority atoms (bottom), giving the spin Seebeck current and the heat current opposite directions.  (b) At unitarity, the scattering rate is inversely proportional to the relative speed.  Therefore, the direction of the spin Seebeck current is reversed relative to (a).}
\label{illustration_seebeck}
\end{figure}

For a polarized gas with a polarization gradient but zero temperature gradient,
where the spin Peltier effect occurs,
the primary current is the spin current which transports minority atoms from the less-polarized region to the more-polarized region.
Since again these minority atoms scatter more often with ``hot'' majority atoms,
the resulting heat current is towards the more-polarized region, resulting in a spin Peltier (heat) current
whose direction is opposite to the primary spin current.
In summary, for a gas far from unitarity, the primary currents and the magnetocaloric currents are in opposite directions.
In other words, the off-diagonal elements in Eq. (\ref{diffusion_Tp}) are negative while the diagonal elements are positive.

(2) At high temperatures ($T \gg T_{F\uparrow}$) and near unitarity ($|k_F a| \gg k_F\lambda > 1$),
the $s$-wave scattering cross section is $1/(k_r/2)^2$ and thus is momentum-dependent,
where $k_r = |{\bf k}_\uparrow - {\bf k}_\downarrow|$ is the relative momentum.
Therefore, the scattering rate is roughly $\sim \frac{1}{k_r^2} \times v_r \sim \frac{1}{k_r}$.
As a result, now minority atoms scatter more often with ``cold'' majority atoms.
Since this is exactly the opposite from the case of far away from unitarity,
the spin Seebeck and spin Peltier currents are reversed relative to the above discussion in (1).
Therefore, at high temperature and unitarity, the primary currents and the magnetocaloric currents are in the same directions,
giving $S_s'$ in Eq. (\ref{diffusion_Tp}) positive sign.
As we will see in the next section, the spin Peltier coefficient $P_s'$ in Eq. (\ref{diffusion_Tp})
is negative for low polarization and becomes positive for high polarization.
This sign change in $P_s'$ comes from the definition of currents and choice of driving forces
and this will be clarified in the section VII where we discuss the Kubo approach.
Directions of the spin Seebeck effect in both limiting regimes are illustrated in figure \ref{illustration_seebeck}.

At low temperatures and unitarity, it is not straightforward to apply the above argument to predict the direction of spin Seebeck and/or spin Peltier currents
since the many body effects may significantly modify the scattering cross section \cite{bruun3, chiacchiera1}, which begins to depend on the center of mass momentum as
well as the relative momentum.
There is, however, a different line of argument that indicates that the sign of the spin Seebeck effect near unitarity remains the same as the temperature is lowered.
Consider low enough temperatures and polarization less than the Chandrasekhar-Clogston limit \cite{clogston, chandrasekhar},
in the superfluid phase \cite{zwierlein, shin}.
When there is a temperature gradient in the system, heat flows ``ballistically'' from the hot region to the cold region
by flow of the normal fluid with respect to the superfluid (in the usual two-fluid description of the superfluid phase).
In the reference frame where the total particle density current vanishes (center of mass frame),
the superfluid flows in the opposite direction to counterbalance the mass current of the normal fluid.
Since the $s$-wave superfluid consists of equal numbers of majority and minority atoms, both the spin current and the heat current are carried only by the normal fluid.
As a result, the spin Seebeck current and the heat current are in the same directions at low temperature in and, presumably, near the superfluid phase.
Therefore, we expect the sign of the spin Seebeck effect to remain the same for all temperatures at unitarity.  Both the thermal conductivity and the spin Seebeck coefficient will diverge at the transition to the superfluid phase.

As another approach to these questions, there is an interesting artificial interaction, namely the Maxwellian interaction where the scattering cross section is
proportional to $1/v_r$ and the Boltzmann equation can be solved exactly in the high temperature limit.
Since the scattering rate does not depend on relative velocity ($(1/v_r) \times v_r =$ constant), there are no spin Seebeck or spin Peltier currents in this case.  Then, if we perturb the scattering cross section around the Maxwellian case, putting in an additional relative-velocity dependence to the scattering rate ``by hand'',
we can perturbatively calculate the spin Seebeck and spin Peltier currents and manipulate the direction of these currents by changing the sign of the perturbation.
This allows us to explicitly show how the magnetocaloric currents are generated from a relative-velocity-dependent cross section.
This will be discussed in the following section in detail.

\section{Quantitative Approach 1 - The Boltzmann Equation}
\subsection{Linearized Boltzmann equation and its scaling}
In the limit of high temperature $T\gg T_F$, the gas is effectively classical, and its
dynamics obey the Boltzmann equation.
In the absence of external forces but with
gradients in local temperature and local densities,
the steady-state Boltzmann equation for species $i$ ($i=\uparrow$ or $\downarrow$) is the following ($i \neq j$):
\begin{align}
&\frac{\hbar}{m}\mathbf{k}_i\cdot \nabla f_i \nonumber\\
&= \int \frac{d^{3}\mathbf{k}_j}{(2\pi)^3} d\sigma \frac{\hbar|\mathbf{k}_{\uparrow} - \mathbf{k}_{\downarrow}|}{m}(f(\mathbf{k}'_{\uparrow})f(\mathbf{k}'_{\downarrow})-f(\mathbf{k}_{\uparrow})f(\mathbf{k}_{\downarrow})),\nonumber\\
\end{align}
where $f(\mathbf{k}_i)$ is the momentum distribution of species $i$,
and the velocity is $\mathbf{v}_{i}=\hbar\mathbf{k}_{i}/m$.
${\bf k}'_\uparrow$ and ${\bf k}'_\downarrow$ are momenta after collision and satisfy the energy momentum conservation.
Working near equilibrium,
we linearize the Boltzmann equation by introducing a small deviation $\psi_i$:
\begin{equation}
f_{i} = f^{0}_{i}(1+ \psi_{i}),
\end{equation}
where $f^{0}_{i}$ is the equilibrium distribution.
In this high temperature regime, the equilibrium distribution is the Boltzmann distribution,
$f^{0}_{i}(k)=n_{i}\lambda^{3}\exp(-E(k)/T)$ and $E(k) = \frac{\hbar^2 k^2}{2m}$.
Since the most relevant length scale in the classical regime is
the thermal de Broglie wavelength $\lambda$,
it is convenient to scale the wave vector {\bf k} with $\lambda$; {\bf k} = ${\bf q}/\lambda$.
Then, $E(q)/T = \frac{q^2}{4\pi}$.

Let's impose the mechanical equilibrium condition. In this classical limit, it is enough to use the ideal gas pressure at equilibrium; $P = nT = (n_{\uparrow}+n_{\downarrow})T$.
From the spatially uniform pressure condition we can relate the density gradients and the temperature gradient via
\begin{equation}\label{uniform_pressure}
\frac{\nabla n}{n} = -\frac{\nabla T}{T},
\end{equation}
to linear order in the gradients.
This relation enables us to express the currents in terms of any two linearly-independent ``driving'' terms such as ($\nabla n_\uparrow, \nabla n_\downarrow$), ($\nabla T, n\nabla p$), ($\nabla T, \nabla(\mu_{\uparrow} - \mu_{\downarrow})$) or any other convenient combinations.
It is convenient to work with ($\nabla n_{\uparrow}$, $\nabla n_\downarrow$) in intermediate stages of the calculation
and then transform it to the desired combination of driving forces using the equilibrium equation of state.
Then the $\lambda$ scaled linearized Boltzmann equations we need to solve become
\begin{align}
&\frac{1}{n_i}\nabla n_i \cdot{\bf q}_i - \frac{1}{n}\left(\frac{q_{i}^{2}}{4\pi} - \frac{3}{2}\right)\nabla n \cdot {\bf q}_i\nonumber\\
&= n_j \int \frac{d^{3}{\bf q}_j}{(2\pi)^3} d\sigma e^{-q^2_j/4\pi}(\psi_{\uparrow}' + \psi_{\downarrow}'-\psi_{\uparrow} - \psi_{\downarrow})|{\bf q}_\uparrow - {\bf q}_\downarrow|~,\label{LinearizedBoltzmannEq}
\end{align}
with $i \neq j$. $\psi'_i$ takes ${\bf q}'_i$ as an argument.

We follow the standard definitions of the particle and energy currents of each species \cite{energy},
\begin{align}
\mathbf{j}_{i} &= \int \frac{d^{3}\mathbf{k}_i}{(2\pi)^3} f_{i}{\bf v}_i =n_i \int \frac{d^{3}{\bf q}_i}{(2\pi)^3}  e^{-q_i^2/4\pi} \psi_{i}\frac{\hbar}{m}\frac{{\bf q}_i}{\lambda}\\
\mathbf{j}_{\epsilon i} &= \int \frac{d^{3}\mathbf{k}_{i}}{(2\pi)^3} f_{i}{\bf v}_i E_i = n_i \frac{T}{4\pi} \int \frac{d^{3}{\bf q}_i}{(2\pi)^3} e^{-q_i^2/4\pi}\psi_{i}\frac{\hbar}{m}\frac{q_i^2 {\bf q}_i}{\lambda} ~.
\end{align}
Because of Galilean invariance, we need to specify an inertial frame.  Except when specified otherwise, we work in the frame
where the local particle current is zero:
${\bf j}_n = {\bf 0}$.

\subsection{Approximate Solution}
The $s$-wave scattering cross section that captures most physics of a short-range interaction is
\begin{equation}\label{general_cross_section}
\frac{d\sigma}{d\Omega} = \frac{a^2}{1 + \left(\frac{k_r a}{2}\right)^2} = \lambda^2 \frac{(a/\lambda)^2}{1 + \left(\frac{q_r a}{2\lambda}\right)^2}~,
\end{equation}
where we scaled the scattering length $a$ with $\lambda$.
An exact solution of the Boltzmann equation for such a cross section is not known and thus we need to resort to approximation methods.
One of the standard ways to find an approximate solution of a linearized Boltzmann equation is the moment expansion method
(for example, see \cite{chiacchiera2}).
Considering all symmetries and assuming the true solution is analytic near small ${\bf q}$,
we take the following ansatz for $\psi_i$ ($i = \uparrow, \downarrow)$:
\begin{equation}\label{ansatz}
\psi_i = -\lambda\sum_{j=\uparrow,\downarrow}\sum_{\ell=0}^LC_{\ell ij}q^{2\ell}_i \frac{\nabla n_{j}\cdot {\bf q}_i}{n_j}~.
\end{equation}
We only consider the case where $\nabla n_\uparrow$ and $\nabla n_\downarrow$ are both parallel to the $z$ axis.
We need to determine the dimensionless coefficients $\{C_{\ell ij}\}$.

The procedure to obtain an approximate solution of the Boltzmann equation is the following:
First, insert the above ansatz into the right hand side of Eq. (\ref{LinearizedBoltzmannEq}).
Then multiply both sides of Eq. (\ref{LinearizedBoltzmannEq}) by
$q^{2\ell}_i q_{iz} \frac{1}{(2\pi)^3}\exp[-q_i^2 /4\pi]$ ($\ell = 0,1,2, ... L$) and integrate out all momenta.
Matching coefficients of density gradients gives $4(L+1)$ linear equations for the $\{C_{\ell ij}\}$,
all of which, however, are not linearly independent
due to Galilean invariance.  We need to fix the reference frame to uniquely determine a solution.
Once we choose an appropriate reference frame (usually ${\bf j_\uparrow} + {\bf j_\downarrow} = 0$),
we have $4(L+1)$ linearly independent equations for the $\{C_{\ell ij}\}$.
Determining the $\{C_{\ell ij}\}$, we have an approximate solution to the Boltzmann equation,
{\it i.e.} an approximate momentum distribution
from which we can calculate all currents of interest.
Here we present results of two limiting cases,
far away from unitarity ($\frac{d\sigma}{d\Omega} = a^2$) and at unitarity ($\frac{d\sigma}{d\Omega} = \frac{4}{k_r^2}$),
which allow an analytic solution (of this approximation) without special functions.
These correspond to the two limits $\lambda/|a|\gg 1$ and $\lambda/|a| \ll 1$, respectively.
For a general scattering length $a$, it is still possible to find a closed form expression
in terms of exponential integrals and incomplete Gamma functions whose arguments depend on $\lambda/|a|$.

Obtaining $C_{\ell ij}$ from a straightforward calculation in the ${\bf j_\uparrow} + {\bf j_\downarrow} = 0$ frame,
we can express the heat current and spin current in terms of $\nabla n_\uparrow$ and $\nabla n_\downarrow$.
Here we choose to express final results in the format of Eq. (\ref{diffusion_Tp})
since we want to study the spin Seebeck coefficient $S_s'$ in detail,
which could be the most directly accessible signature of the spin Seebeck effect in experiments.
Therefore, we transform these two gradients to $\nabla T$ and
$n\nabla p$,
using the equilibrium equation of state and the mechanical equilibrium condition.
The transformation matrix is
\begin{align}
\begin{pmatrix}
\nabla n_\uparrow \\
\nabla n_\downarrow
\end{pmatrix}
=
\begin{pmatrix}
-\frac{T}{n} && -\frac{T}{n}\\
\frac{2n_\downarrow}{n} && -\frac{2n_\uparrow}{n}
\end{pmatrix}
^{-1}
\begin{pmatrix}
\nabla T \\
n\nabla p
\end{pmatrix}~.
\end{align}

Since the left hand side of Eq. (\ref{LinearizedBoltzmannEq}) is a third-order polynomial in $q$,
the simplest ansatz is with $L = 1$.
In principle, we can go up to any order in $L$ we want, but an $L \geq 2$ ansatz complicates the computation,
while the simplest ansatz already exhibits nontrivial results.
Furthermore, we find that the change in $S_s'$ on moving from the $L=1$ to the $L=2$ approximation is quite small:
about a 1\% change near unitarity and in the value of $\lambda/|a|$ at the zero crossing, growing to near 7\%
far from unitarity.
Therefore, here we only present results of $L = 1$.

\begin{widetext}
We present our results in the conventional format (without $\lambda$ scaling): Near unitarity ($|a|\gg\lambda$),
\begin{equation}\label{unitarity_zeromass}
\begin{pmatrix}
\mathbf{j}_{heat}\\
\mathbf{j}_{spin}
\end{pmatrix}
=-\frac{45\pi^{3/2}}{608\sqrt{2}}\left(\frac{\hbar}{m}\left(\frac{T}{T_{F}}\right)^{3/2}\right)
\begin{pmatrix}
\kappa_u' && \frac{1}{2}\frac{(n_\uparrow - n_\downarrow)T n}{n_\uparrow n_\downarrow} - \frac{39}{10}T\ln \left(\frac{n_\uparrow}{n_\downarrow}\right) \\
\frac{(n_\uparrow - n_\downarrow)}{T}  && \frac{39}{10}
\end{pmatrix}
\begin{pmatrix}
\nabla T\\
n\nabla p
\end{pmatrix}~,
\end{equation}
where $\kappa_u'$ (proportional to the thermal conductivity at unitarity) is
\begin{equation}\label{kappa_u}
\kappa_u' = \frac{5}{16}n\left(\frac{(73n_{\uparrow}^{2} + 82n_{\uparrow}n_{\downarrow} + 73n_{\downarrow}^{2})}{n_{\uparrow}n_{\downarrow}}\right) -(n_\uparrow - n_\downarrow)\ln\left(\frac{n_\uparrow}{n_\downarrow}\right).
\end{equation}
Far away from unitarity ($|a| \ll \lambda$),
\begin{equation}\label{away_zeromass}
\begin{pmatrix}
\mathbf{j}_{heat}\\
\mathbf{j}_{spin}
\end{pmatrix}
=-\frac{15\pi^{3/2}}{224\sqrt{2}}\left(\frac{\hbar}{m (k_{F}a)^{2}}\sqrt{\frac{T}{T_{F}}}\right)
\begin{pmatrix}
\kappa_a' && -\frac{1}{2}\frac{(n_\uparrow - n_\downarrow) Tn}{n_\uparrow n_\downarrow} - \frac{43}{10}T\ln\left(\frac{n_\uparrow}{n_\downarrow}\right) \\
-\frac{(n_\uparrow - n_\downarrow)}{T} && \frac{43}{10}
\end{pmatrix}
\begin{pmatrix}
\nabla T\\
n\nabla p
\end{pmatrix}~,
\end{equation}
where $\kappa_a'$ (proportional to the thermal conductivity far away from unitarity) is
\begin{equation}\label{kappa_a}
\kappa_a' = \frac{5}{8}n\left(\frac{(29n_{\uparrow}^{2} + 26n_{\uparrow}n_{\downarrow} + 29n_{\downarrow}^{2})}{n_{\uparrow}n_{\downarrow}}\right) + (n_\uparrow - n_\downarrow)\ln\left(\frac{n_\uparrow}{n_\downarrow}\right).
\end{equation}
\end{widetext}

\begin{figure}
\includegraphics[width=3.50in]{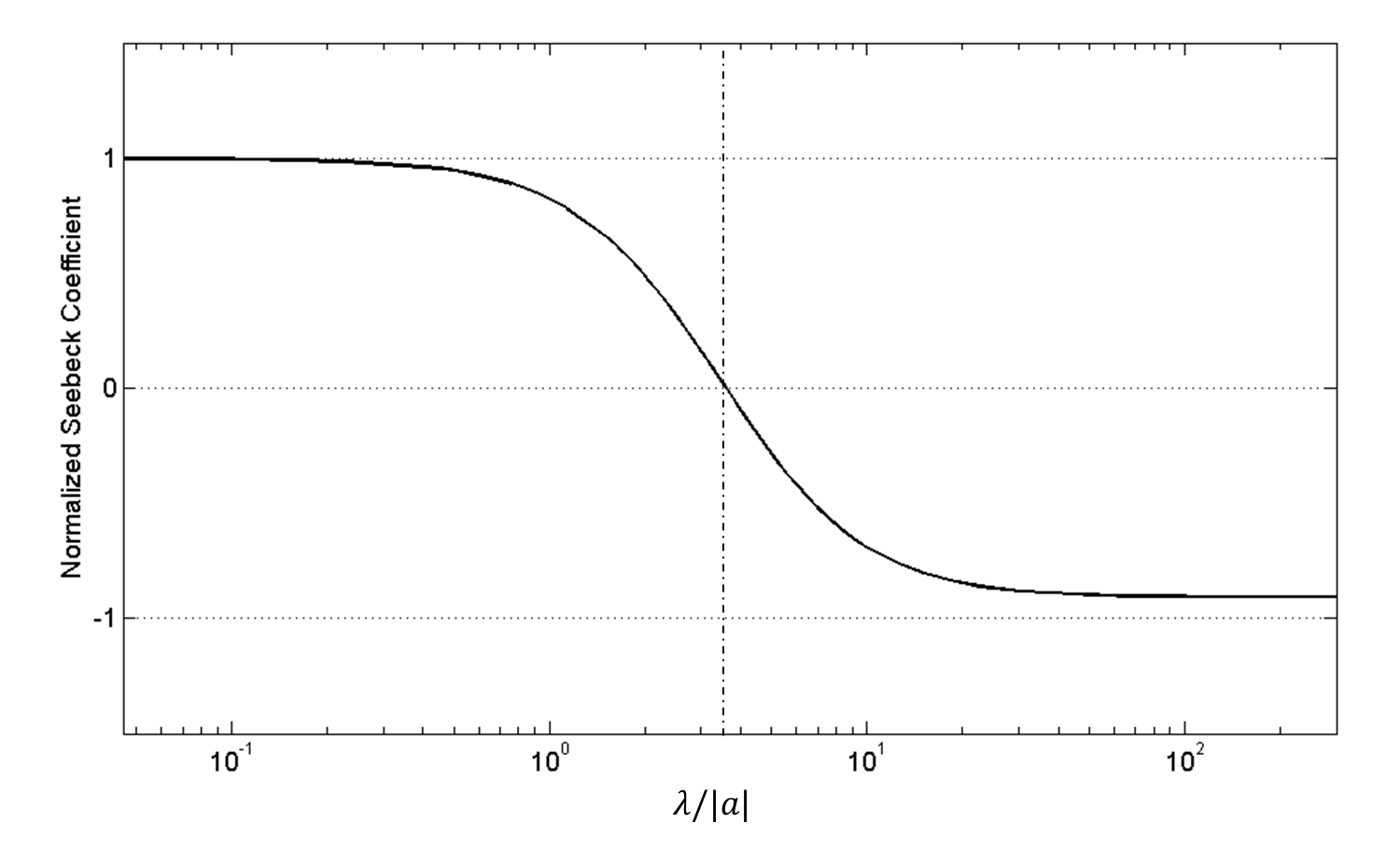}
\centering
\caption{Normalized Seebeck coefficient $S_s'$ as a function of $\lambda / |a|$ in the classical regime. The normalization is by a factor of $\frac{15}{608\sqrt{2}}\left(\frac{\hbar}{m}\frac{1}{T \lambda^3}\frac{1+ 4\pi(a/\lambda)^2}{(a/\lambda)^2}\right) \frac{(n_\uparrow - n_\downarrow)}{n}$. This choice of normalization, which is inversely proportional to a typical value of scattering cross section with scaled scattering length $a/\lambda$, gives finite values at both limits and gives the Seebeck coefficient $1$ at unitarity (Eq. (\ref{unitarity_zeromass})). We can see that $S_s'$ changes sign near $\lambda / |a| \simeq 3.62$. Since we factored out all explicit temperature and polarization dependence in the normalization choice and scaled the scattering length, this plot remains the same for all temperature and polarization ranges at this order of approximation ($L=1$). }
\label{scatteringlength_seebeck}
\end{figure}

The above matrices clearly exhibit the existence of the spin Seebeck and spin Peltier effects (non-vanishing off diagonal terms)
only for nonzero polarization.  We will mostly focus now on the spin Seebeck coefficient $S_s'$, which gives the spin current
due to a temperature gradient in the absence of a spin polarization gradient.

As argued in the previous section,
the spin Seebeck coefficient changes sign as a function of interaction strength.
Near unitarity (Eq. \ref{unitarity_zeromass}), it is positive so the spin Seebeck current and the heat current are in the same direction.
Far away from unitarity (Eq. \ref{away_zeromass}), it is negative so the spin Seebeck current and the heat current are in the opposite direction.

Next let's consider the polarization dependence of the transport coefficients.
The heat current and the temperature gradient are even under spin index exchange
while the spin current and the polarization gradient are odd.
Therefore, thermal conductivity and spin diffusivity are
even functions of polarization while magnetocaloric effects are odd functions of polarization.
The above matrices satisfy these polarization parity constraints and
the form of the polarization dependence of the transport coefficients is
the same in both limits of large and small $\lambda/|a|$.
In fact, we can show that the polarization dependence
(thus $n_\uparrow$ and $n_\downarrow$ dependence) of the transport coefficients
maintains this form for all values of $\lambda/|a|$ and to all orders of approximation.
The proof is given in Appendix A.
Here we study the Seebeck coefficient $S_s'$ in detail, which is our prime interest.

Although it is conventional to scale the scattering length $a$ with $k_F$ (as we did in the above matrices),
it is easier to see the structure of the Seebeck coefficient in terms of $\lambda / |a|$ in classical regime.
Once we obtain an approximate solution of the Boltzmann equation with a general $\lambda/|a|$ and $L = 1$,
we can explicitly show that
\begin{align}\label{scaled_seebeck}
S_s' = \frac{\hbar}{m}\frac{n_\uparrow - n_\downarrow}{n \lambda^3}\frac{1}{T}h_1(\lambda / |a|)~,
\end{align}
where $h_1(x)$ is a dimensionless function that contains $Ei(x)$, the exponential integral,
and diverges as $\lambda/|a|\rightarrow \infty$ (see Eq. (\ref{away_zeromass})).
Since it contains no explicit temperature or polarization dependence,
$h_1(x)$ is independent of polarization and temperature at this order of approximation ($L=1$).
Therefore, once we scale the scattering length by $\lambda$ and factor out dimensionful parameters and polarization,
the dependence of $S_s'$ on the scattering length is determined by $h_1(x)$ and
the value of $\lambda / |a|$ at which $S_s'$ crosses zero is solely determined by the equation
$h_1(x) = 0$, which is independent of temperature and polarization.
In the $L=1$ approximation, the zero-crossing value is $\lambda/|a| \simeq 3.62$.
Figure \ref{scatteringlength_seebeck} is a plot of the normalized $S_s'$ as a function of $\lambda/|a|$.
In the $L=2$ approximation, we find that the scaling function $h_1(x)$ slightly changes to $h_2(x)$
and the zero-crossing point remains at $\lambda/|a|\simeq 3.62$.
In Appendix A, we show that the structure of Eq. (\ref{scaled_seebeck})
(and other transport coefficients in a similar manner) remains
to all orders of approximation.
Therefore, we may conclude that in this classical regime the Seebeck coefficient is
linearly proportional to the polarization $p$ and inversely proportional to $T \lambda^3$,
once we scale the scattering length by $\lambda$.

Note that Eqs. (\ref{unitarity_zeromass}) and (\ref{away_zeromass}) do not explicitly satisfy the Onsager relation
and the spin Peltier coefficient $P_s'$ is still negative for low polarization even near unitarity.
These come from the definition of diffusive currents and choice of representation
and will be discussed in detail in the next section in terms of the Kubo formula.
For now, we will focus on the spin Seebeck coefficient $S_s'$
which appears to be the most promising candidate of the magnetocaloric effects to be detected in experiments.

From the Einstein relation, we obtain the thermal diffusivity $D_T$
after dividing $\kappa_u'$ and $\kappa_a'$ by $C_P = 5n/2$, the heat capacity per volume at fixed pressure and polarization.
These results confirm the ``power-counting'' estimates of diffusivities:
In case of an unpolarized gas ($n_{\uparrow} = n_{\downarrow}$),
$D_s$ at unitarity is $\cong 1.1 \frac{\hbar}{m}\left(\frac{T}{T_F}\right)^{3/2}$,
which is consistent with previous work \cite{sommer1, bruun1}.
Also, the thermal diffusivity does satisfy the inequality, $D_T > D_s$ at zero polarization.
Furthermore, for a highly polarized gas ($n_{\uparrow} \gg n_{\downarrow}$),
$D_T \sim \frac{n^{2}}{n_{\uparrow}n_{\downarrow}}D_s \sim \frac{n_{\uparrow}}{n_{\downarrow}}D_s$,
which is also as expected from the ``power-counting'' estimates.

\subsection{Estimate of the Spin Seebeck Effect}
The spin Seebeck effect seems to be more accessible to experiment than the spin Peltier effect,
since the measurement of spin currents has already been done \cite{sommer1,sommer2} and seems more straightforward than measuring heat currents.
Also, the spin mode diffuses slower than the heat mode, so the change in spin polarization produced by the spin Seebeck effect will relax slowly, enhancing its detectability.
An initially fully equilibrated spin-polarized gas could be heated at one end,
producing a temperature gradient, and then the resulting spin current could be measured if it is large enough.

\begin{figure}
\includegraphics[width=3.50in]{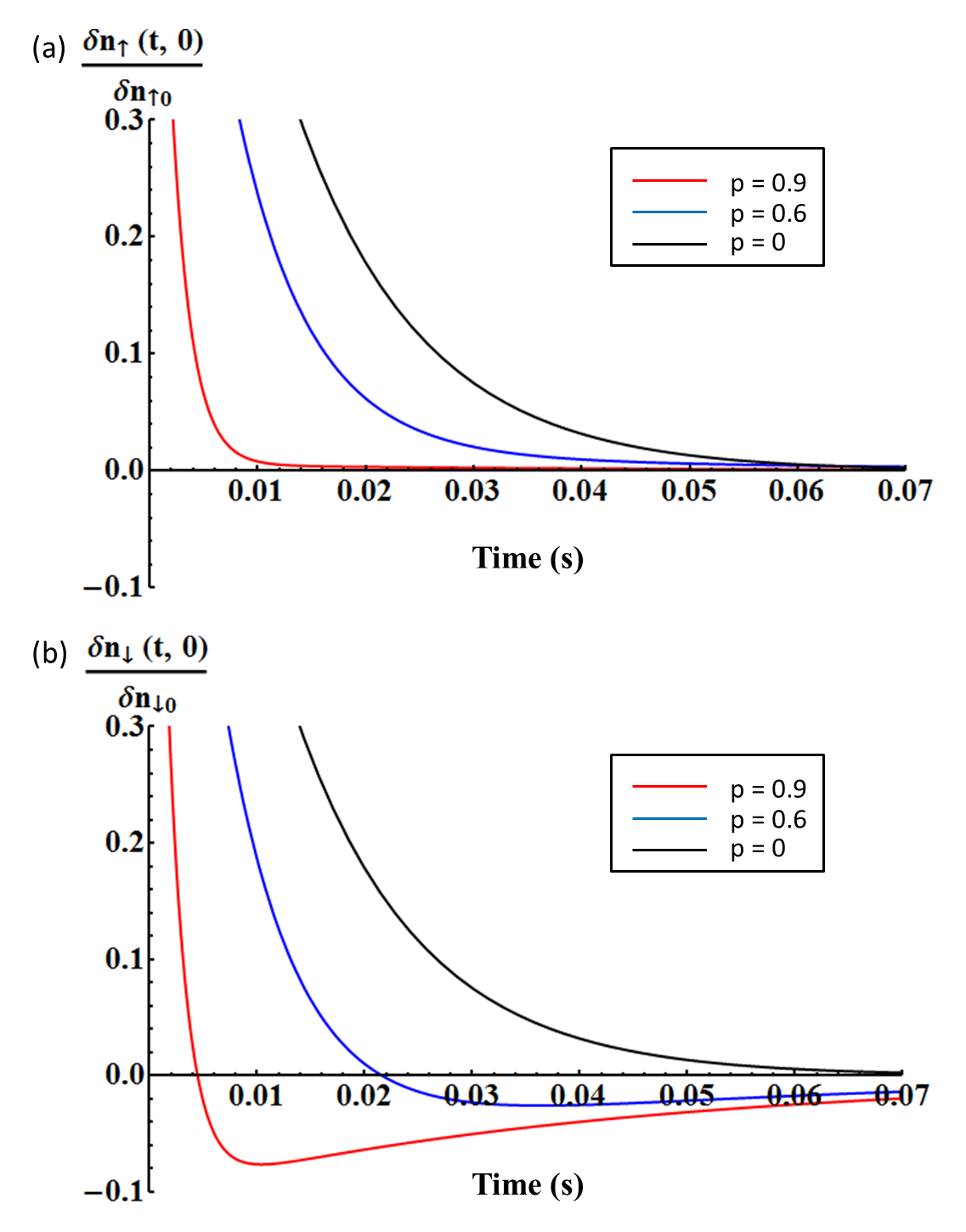}
\centering
\caption{(Color online) $\delta n_i(t,0)$ normalized by initial deviation $\delta n_{i0}$ as a function of time near unitarity.
(a) Majority density deviation relaxes monotonically for any polarization. (b) Minority density deviation shows nonmonotonic relaxation for $p>0$ due to the spin Seebeck effect.
Here we assume $^6$Li atoms with $T/T_F = 4$ and the longitudinal length of the trap $L = 200 {\rm \mu m}$ to set the time scale.
}
\label{deviation_unitarity}
\end{figure}

\begin{figure}
\includegraphics[width=3.50in]{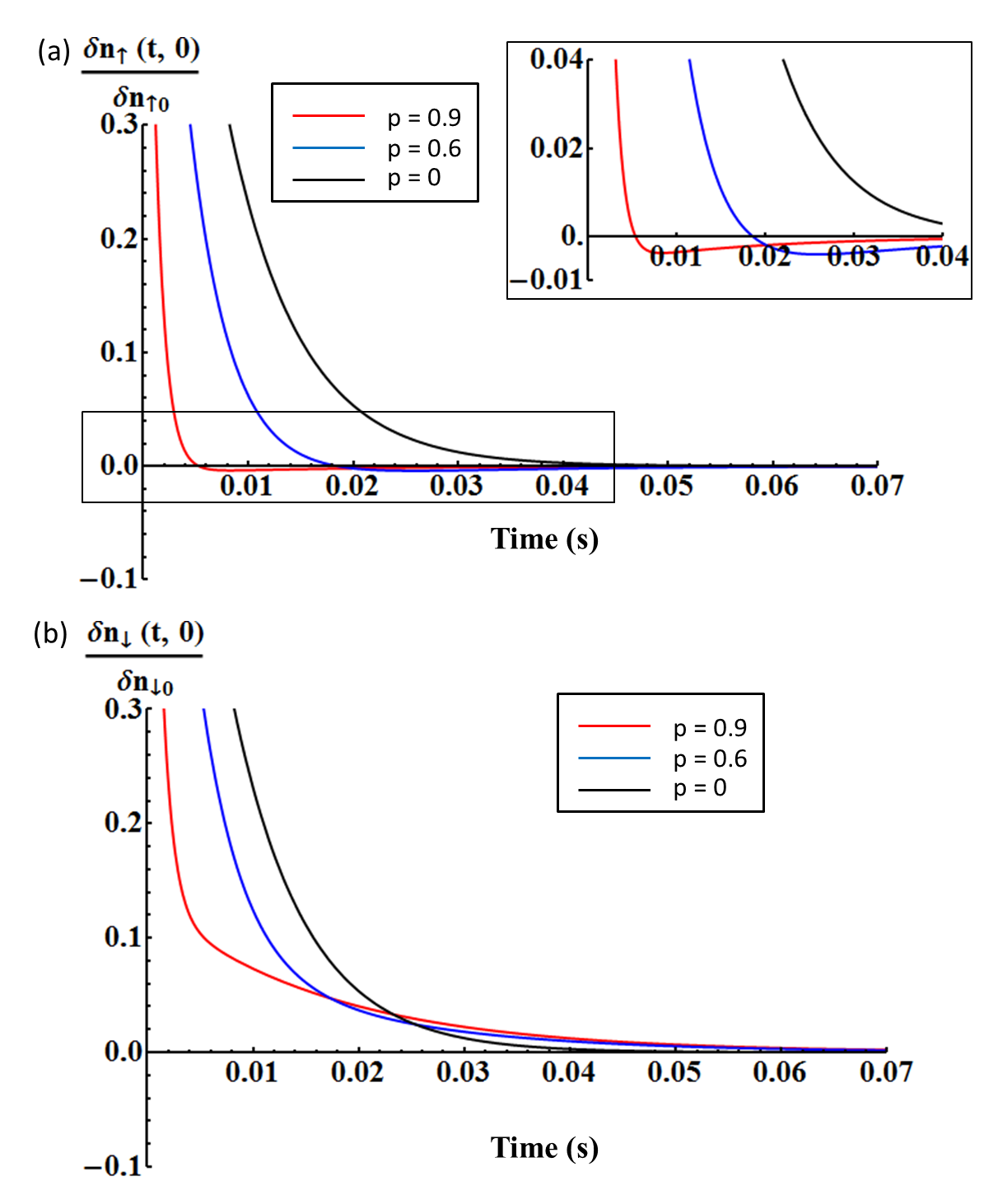}
\centering
\caption{(Color online) $\delta n_i(t,0)$ normalized by initial deviation $\delta n_{i0}$ as a function of time far away from unitarity.
(a) Majority density deviation relaxes nonmonotonically for $p>0$ due to the spin Seebeck effect. Inset figure magnifies the nonmonotonic part of majority density deviations which are very weak compared to the minority density deviations at unitarity shown in Fig. 3.  (b) Minority density deviation relaxes monotonically for any polarization. Here we assume $^6$Li atoms with $T/T_F = 4$ and $\lambda/|a| = 4 $ and the longitudinal length of the trap $L = 200 {\rm \mu m}$ to set the time scale.}
\label{deviation_away}
\end{figure}

Let's make quantitative estimates of signatures of the spin Seebeck effect that are relevant to such a proposed experiment.
We will make our estimates for a gas in a uniform potential, but the results should be roughly correct for a gas cloud in a trap if one compares points at opposite ends of the cloud that are at the same potential so will have the same local densities and polarization at equilibrium.
First, apply a small, long wavelength temperature inhomogeneity along the $z$ axis.
In mechanical equilibrium, nonuniform temperature implies nonuniform total density (by Eq. (\ref{uniform_pressure})),
thus temperature modulation implies density modulation. This enables us to write the initial total density as
\begin{equation}
n(t=0, z) = n_{0} + \delta n_{0}\cos wz,
\end{equation}
where $w$ ($= \pi/L$) is the wavenumber of the modulation, $L$ is the length of the system over which the full temperature difference is applied,
and $\delta n_{0}$ is the small deviation of total density from the mean value $n_{0}$ due to this temperature difference (at uniform pressure).
Eq. (\ref{uniform_pressure}) implies that if we initially locally heat $z = L$ relative to $z=0$,
then this location initially has lower density because of thermal expansion.
From the initial condition of uniform polarization,
the density of each spin component at $t = 0$ is
\begin{align}
n_{i}(t=0,z) &= n_{i 0} + \delta n_{i 0}\cos wz\\
\delta n_{i 0} &= \frac{n_{i 0}}{n_{0}}\delta n_{0}.
\end{align}
Then, we write a diffusion matrix in the form of Eq.(\ref{diffusion}), assuming that there are only these diffusive currents (no initial other motion of the gas).

As mentioned above in Section III, the diffusion matrix has two eigenmodes,
the spin mode with the eigenvalue ${\frak D}_s$ and the thermal mode with the eigenvalue ${\frak D}_T$.
Applying the continuity equation to the diffusion matrix, we obtain a coupled set of heat equations with appropriate initial conditions, which can be solved immediately for $t > 0$ in terms of the two eigenmodes.
The solution of the heat equation gives us the space-time dependence of density of each species:
\begin{align}
\begin{pmatrix}
n_{\uparrow}(t,z)\\
n_{\downarrow}(t,z)
\end{pmatrix}
&=
\begin{pmatrix}
n_{\uparrow 0}\\
n_{\downarrow 0}
\end{pmatrix}
+\alpha e^{-{\frak D}_{s}w^{2}t}\delta n_{\downarrow 0}
\begin{pmatrix}
\gamma \\
1
\end{pmatrix}
\cos wz
\nonumber\\
&+
\beta e^{-{\frak D}_{T}w^{2}t}\delta n_{\downarrow 0}
\begin{pmatrix}
\zeta \\
1
\end{pmatrix}
\cos wz ~,
\end{align}
where $\alpha, \beta, \gamma, $ and $\zeta$ are determined by initial conditions. See Appendix B for details.
Since ${\frak D}_T > {\frak D}_s >0$, the two diffusive modes relax to global equilibrium at different rates.
At nonzero polarization, $|\gamma|, |\zeta| \neq 1$.
From these two properties,
the density deviations of each species have different time evolution from one another:
the density deviation of one species relaxes nonmonotonically
while the density of the other species relaxes monotonically.
This nonmonotonic relaxation of density deviation is a qualitative signature of the spin Seebeck effect.
Near unitarity, we already know that the heat current and the spin current due to a temperature gradient are in the same direction.
Therefore, the initially colder region ($z < \pi/2w$) becomes more polarized due to the spin Seebeck current.
It turns out that the minority density deviation changes sign as it relaxes towards equilibrium.
Far from unitarity, on the other hand, the initially cold region becomes less polarized,
since the direction of the spin current is reversed, resulting in nonmonotonic relaxation of the majority density deviation.
Figures \ref{deviation_unitarity} and \ref{deviation_away} are plots of
relaxation of the density deviations of each species as a function of time for the two cases of near unitarity and of far away from unitarity.

\begin{figure}
\includegraphics[width=3.50in]{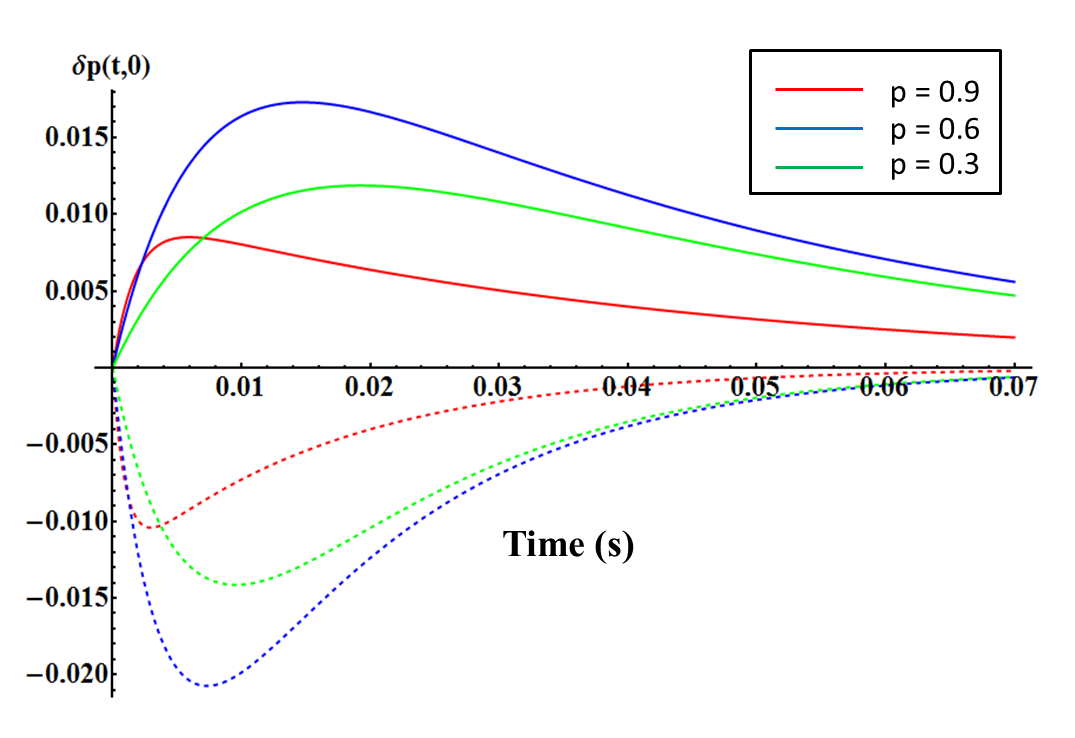}
\centering
\caption{(Color online)) Normalized polarization deviation as a
function of time for three global polarization values, p = 0.9 (lower
curve), 0.6 (highest curve), and 0.3 (center curve). Near unitarity (solid lines) the polarization deviation is positive while far away from unitarity (dashed lines) it is negative. We found that the polarization deviation is the largest near $p=0.6$ in both limits. Same physical parameters as Figures \ref{deviation_unitarity} and \ref{deviation_away} were used to fix the time scale. The polarization deviation $\delta p (t,0)$ at the initially cold end of the cloud is normalized by the initial total density deviation, $\delta n_0 / n_0$.}
\label{polarization}
\end{figure}

Another consequence of the spin Seebeck effect is the change in polarization.
As argued, the sign of the local polarization change depends on the direction of the spin Seebeck current.
Figure \ref{polarization} shows the (normalized) deviation of polarization from the average value as a function of time
at $z$ = 0.
As expected, deviation is positive near unitarity and is negative far away from unitarity.

\begin{figure}
\includegraphics[width=3.00in]{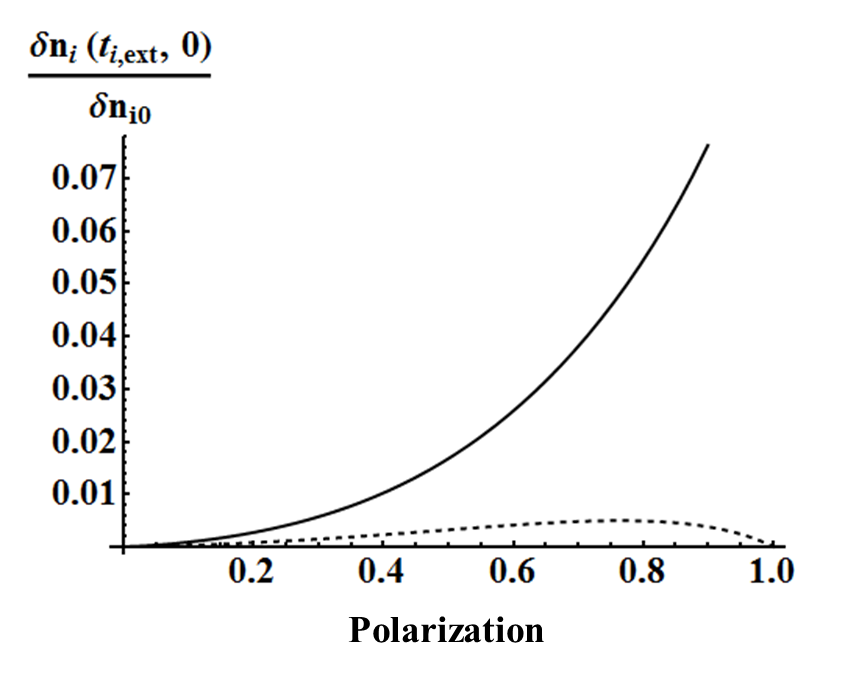}
\centering
\caption{Extremum density deviation normalized by initial deviation, $\left|\frac{\delta n_{i}(t_{i,ext},0)}{\delta n_{i0}}\right|$ as a function of polarization.
Near unitarity (solid line), $i = \downarrow$; far from unitarity (dashed line), $i = \uparrow$. $t_{i,ext}$ is the time when the density deviation of species $i$
reaches its extremum.
Far from unitarity (dashed line), the normalized extremum density deviation vanishes in the high polarization limit since there are no minority atoms to scatter from.}
\label{delta_n_ext}
\end{figure}

Perhaps one of the most easily accessible quantities in experiment
is the density deviation of the species that shows nonmonotonic relaxation vs. time.
A dimensionless measure of the extremum deviation is
$\left|\frac{\delta n_{i}(t_{i,ext},0)}{\delta n_{i0}}\right|$,
where $i = \uparrow$ away from unitarity and $i = \downarrow$ near unitarity
and $t_{i,ext}$ is the time when density deviation of species $i$ is its extremum of opposite sign from the initial condition.
Figure \ref{delta_n_ext} shows the dimensionless measure of the extremum density deviation as a function of polarization.
Near unitarity (solid line), it is a monotonically increasing function of polarization for $p<1$
(at $p=1$, $n_\downarrow$ is zero so the deviation of $n_\downarrow$ is undefined).
Around $p \approx 0.8$, the strength of the spin Seebeck effect by this measure is about 5 $\%$.
Far from unitarity (dashed line), this signal is much weaker.
This is expected since the spin Seebeck effect is a consequence of interaction
and does not exist in the non-interacting gas.
Thus, the spin Seebeck effect should be strongest near unitarity.
Note that in both limits, the spin Seebeck effect disappears when the gas is unpolarized, $p = 0$.

The spin Seebeck effect is a small effect.  In the regimes where we have been able to estimate it and using the measures we have been able to devise,
it is less than a $10\%$ effect.
However, it is worth emphasizing that these computations are done in the classical regime, so
the spin Seebeck effect does not demand extremely low temperatures to detect it.
We expect it to be most readily detected at temperatures of order $T_F$,
where the diffusivities are minimized so the resulting time scales are longest.
In addition, the procedure to detect it discussed in this section does not require knowledge of the system's equation of state.
Importantly, we propose a type of experiment where the the spin Seebeck effect is a {\it qualitative effect}, namely a nonmonotonicity of the system's relaxation to global equilibrium.

\subsection{Exactly Solvable Model}
So far, our approach was based on an approximation method.
Therefore, it is worthwhile to compare the main findings to a different approach, namely perturbing around the Maxwellian model,
where the scattering cross section is inversely proportional of the relative speed \cite{smith}.
The linearized Boltzmann equation can be exactly solved for the classical Maxwellian model.

Let's recall the argument from which we determined directions of the magnetocaloric effects.
Since the scattering rate is proportional to the product of the cross section and relative speed,
the scattering rate is independent of momentum for the
Maxwellian interaction.
Therefore, we expect that $S_s' = 0$ for any polarization.
It is straightforward to exactly solve the Boltzmann equation using the ansatz Eq. (\ref{ansatz}) with $L = 1$ to confirm this.

The next step is to
perturb the Maxwellian scattering cross section to generate magnetocaloric currents.
One simple way to perturb the cross section is to add a small term
which depends linearly on relative momentum to the original cross section:
\begin{equation}\label{maxwellian}
\frac{d\sigma}{d\Omega} = S_{0}\left(\frac{1}{k_{r}} + \epsilon k_{r}\right),
\end{equation}
where $S_{0}$ is a constant of the dimension of length and $\epsilon$ is a small expansion parameter.
For a positive $\epsilon$, the momentum dependence of the scattering rate is similar to the case away from unitarity,
higher scattering rate for higher relative momentum.
Thus we expect the resulting spin Seebeck current is in the opposite direction from the primary heat current.
For a negative $\epsilon$, the momentum dependence resembles the case near unitarity and therefore
we expect the resulting spin Seebeck current is in the same direction as the primary heat current.

To find the solution of the Boltzmann equation up to linear order in $\epsilon$,
we need $L = 2$ in the ansatz of Eq. (\ref{ansatz}).
After a straightforward calculation, we obtain the following results:
\begin{widetext}
\begin{align}\label{maxwell_zeromass}
\begin{pmatrix}
\mathbf{j}_{heat}\\
\mathbf{j}_{spin}
\end{pmatrix}
=-D_{0}
\begin{pmatrix}
\kappa_M'
&& -\frac{\pi\epsilon (n_{\uparrow} - n_{\downarrow})Tn}{ 2n_\uparrow n_\downarrow \lambda^{2} } - \frac{T}{10}\left(1 - \frac{20\pi\epsilon}{\lambda^{2}}\right)\ln\left(\frac{n_\uparrow}{n_\downarrow}\right)\\
-\frac{\pi\epsilon (n_{\uparrow} - n_{\downarrow})}{\lambda^{2} T}  && \frac{1}{10}\left(1 - \frac{20\pi\epsilon}{\lambda^{2}}\right)
\end{pmatrix}
\begin{pmatrix}
\nabla T \\
n\nabla p
\end{pmatrix}~,
\end{align}
where $\kappa_M'$ (proportional to the thermal conductivity of the Maxwellian Model) is
\begin{align}\label{kappa_m}
\kappa_M' = \frac{1}{2}n\left(\frac{(n_{\uparrow}^{2} + n_{\uparrow}n_{\downarrow} + n_{\downarrow}^{2})}{n_{\uparrow}n_{\downarrow}} - \frac{3\pi\epsilon}{\lambda^{2}}\frac{(9n_{\uparrow}^{2} + 10n_{\uparrow}n_{\downarrow} + 9n_{\downarrow}^{2})}{n_{\uparrow}n_{\downarrow}}\right) + \frac{\pi\epsilon (n_\uparrow - n_\downarrow)}{\lambda^2} \ln\left(\frac{n_\uparrow}{n_\downarrow}\right)~.
\end{align}
\end{widetext}

$D_0$ is $\frac{5\hbar}{mS_{0}n\lambda^{2}}$, with units of a diffusivity.

We immediately see that all off-diagonal elements vanish for zero polarization.
When $\epsilon = 0$, we see that $S_s' = 0$
and thus we conclude that the spin Seebeck effect is a consequence of momentum dependence of the scattering rate.
The sign of the spin Seebeck current at nonzero $\epsilon$ is as expected.
For $P_s'$, we again see the logarithmic term which
will be discussed in the following section.
Perturbing the Maxwellian scattering cross section
re-confirms the sign argument for the spin Seebeck effect that we presented in the previous Section.

\section{Structure of Transport Coefficients - Kubo Formula}
The Kubo formula gives a formally exact expression for the transport coefficients in the linear response regime (see e.g. \cite{mori,luttinger}).
For irreversible processes in the linear response regime,
what the the Kubo formula gives are the transport coefficients for the dissipative forces and currents associated with the entropy production.
At mechanical equilibrium for our two-species gas,
the dissipative forces are $\nabla T$ and $\nabla (\mu_\uparrow - \mu_\downarrow)$ \cite{chaikin},
thus what we obtain from the Kubo formula is the diffusion matrix in the form of Eq. (\ref{diffusion_Tmu}),
whose off-diagonal terms always satisfy the Onsager relation.
Thus, the diffusion matrix in the form of Eq. (\ref{diffusion_Tp}),
in which we summarized the results in the previous section,
does not generally satisfy the Onsager relation, although it is possibly easier to observe experimentally.
In Appendix C, we summarize the results in the form of Eq. (\ref{diffusion_Tmu})
and explicitly show both our approximate solutions and the perturbative solution from the exactly solvable Maxwellian model
indeed satisfy the Onsager relation.

Following Ref. \cite{mori} and from Eq. (\ref{diffusion_Tmu}), the spin Seebeck coefficient $S_s$ and Peltier coefficient $P_s$
can be expressed via the Kubo formula as
\begin{align}\label{seebeck_kubo_formal}
T S_s &=\frac{1}{3VT}\int^{\infty}_0 dt \langle{\bf J}_{heat}(0)\cdot{\bf J}_{spin}(t)\rangle \\
&=\frac{1}{3VT}\int^{\infty}_0 dt \langle{\bf J}_{heat}(t)\cdot{\bf J}_{spin}(0)\rangle = P_s,
\end{align}
where $V$ is the total volume of the system
and the current ${\bf J}$ is the volume integral of local current density,
\begin{equation}
{\bf J}(t) = \int d^3 {\bf x} \;  {\bf j}({\bf x},t).
\end{equation}
The average is taken over the equilibrium distribution,
which is just the Boltzmann distribution of each species in the high-temperature classical regime.

Since we assume Galilean invariance, the total momentum of the entire gas is conserved.
Therefore, any physical quantity which is transported with the total particle current
${\bf J}_n$ remains finite for all time $t$ and gives a divergent contribution in the Kubo formula, meaning
that quantity moves ``ballistically'' rather than diffusively.
Therefore, it is crucial to use the frame-independent definition of the local diffusive currents from Eq. (\ref{heat}) and Eq. (\ref{spin}).
Let's write them again and slightly manipulate the heat current:
\begin{align}
{\bf j}_{spin} &= \frac{n_\downarrow}{n}{\bf j}_\uparrow - \frac{n_\uparrow}{n}{\bf j}_\downarrow\\
{\bf j}_{heat} &= {\bf j}_{\epsilon} - \mu_\uparrow {\bf j}_\uparrow - \mu_\downarrow {\bf j}_\downarrow - s T {\bf j}_n \nonumber\\
&= {\bf j}_{\epsilon} - \frac{5}{2}T{\bf j}_n - (\mu_\uparrow - \mu_\downarrow){\bf j}_{spin}\label{heat_rephrase},
\end{align}
where $s T = \frac{5}{2}T - \frac{n_\uparrow}{n}\mu_\uparrow - \frac{n_\downarrow}{n}\mu_\downarrow$ and ${\bf j}_{\epsilon} = {\bf j}_{\epsilon\uparrow} + {\bf j}_{\epsilon\downarrow}$.
Defining ${\tilde {\bf j}}_{heat} \equiv {\bf j}_\epsilon - (5T/2) {\bf j}_n$ \cite{energy_current}, the Kubo formula gives us the following spin Seebeck coefficient:
\begin{align}
TS_s = P_s &= \frac{1}{3VT} \int^{\infty}_0 dt [ \langle{\tilde {\bf J}}_{heat}(0)\cdot{\bf J}_{spin}(t)\rangle \nonumber\\
&\quad\quad - (\mu_\uparrow - \mu_\downarrow)\langle{\bf J}_{spin}(0)\cdot{\bf J}_{spin}(t)\rangle]\\
&= \frac{1}{3VT} \int^{\infty}_0 dt \langle{\tilde {\bf J}}_{heat}(0)\cdot{\bf J}_{spin}(t)\rangle \nonumber\\
&\quad- T\ln\left(\frac{n_\uparrow}{n_\downarrow}\right)\sigma_s \label{seebeck_kubo_log},
\end{align}
where $\sigma_s$ is the spin conductivity given in Eq. (\ref{diffusion_Tmu}) and $\mu_\uparrow - \mu_\downarrow = T \ln(n_\uparrow/n_\downarrow)$.
Therefore, in this representation,
the spin Seebeck coefficient (and thus also the spin Peltier coefficient)
always carries an additional term of the spin conductivity $\sigma_s$ multiplied by $-\ln(n_\uparrow/n_\downarrow)$.
This is the origin of that term in $P_s'$ in Eqs. (\ref{unitarity_zeromass}), (\ref{away_zeromass}), and (\ref{maxwell_zeromass}).
We can understand the reason why the spin Seebeck coefficient $S_s'$ does not include such a term
from the following observation:
At mechanical equilibrium and high temperature, we have
\begin{align}\label{mu_transform}
\nabla (\mu_\uparrow - \mu_\downarrow) = \ln\left(\frac{n_\uparrow}{n_\downarrow}\right)\nabla T + T \frac{n^2}{2n_\uparrow n_\downarrow}\nabla p~.
\end{align}
Therefore, that $\nabla p = 0$ implies
$\nabla (\mu_\uparrow - \mu_\downarrow) = \ln(n_\uparrow/n_\downarrow)\nabla T$.
The spin current now becomes
\begin{align}
{\bf j}_{spin} &= -\sigma_s \nabla(\mu_\uparrow - \mu_\downarrow) - S_s \nabla T \\
&= - D_s n\nabla p - \left(\sigma_s \ln(n_\uparrow/n_\downarrow) + S_s\right)\nabla T \\
&= - D_s n\nabla p - S_s' \nabla T~.
\end{align}
Thus the additional $\sim \ln(n_\uparrow/n_\downarrow)$ term from Eq. (\ref{mu_transform})
exactly cancels the similar term from Eq. (\ref{seebeck_kubo_log}).
Consequently, what we have computed for $S_s'$ in the previous section is
the first term in Eq. (\ref{seebeck_kubo_log}) to which the sign argument in section IV should be applied.

We can extract more information from Eq. (\ref{heat_rephrase}).
In ${\bf j}_n = 0$ frame, the heat current is
${\bf j}_{heat} = {\bf j}_{\epsilon} - T\ln(n_\uparrow/n_\downarrow){\bf j}_{spin}$.
Thus, heat conductivity and spin Peltier coefficient always have
contributions coming from spin current. Therefore, in the format of Eq. (\ref{diffusion_Tp})
we can write
\begin{align}
\kappa' &= \kappa'_1 - T\ln(n_\uparrow/n_\downarrow) S_s' \\
P_s' &= P_{s1}' - T\ln(n_\uparrow/n_\downarrow) D_s ~,
\end{align}
where $\kappa'_1$ and $P_{s1}'$ are first terms which do not originate from spin current.
Results in the previous section clearly show this.

Lastly, we study the thermal conductivity, $\kappa$, in Eq. (\ref{diffusion_Tmu}).
\begin{align}
\kappa &= \frac{1}{3V T^2}\int_0^\infty dt <{\bf J}_{heat}(t)\cdot{\bf J}_{heat}(0)> \\
&=\frac{1}{3 V T^2}\int_0^\infty dt <{\tilde {\bf J}}_{heat}(t)\cdot{\tilde{\bf J}}_{heat}(0)> \nonumber\\
&\;\;- 2T\ln\left(\frac{n_\uparrow}{n_\downarrow}\right)S_s' + T \left(\ln\left(\frac{n_\uparrow}{n_\downarrow}\right)\right)^2 \sigma_s ~. \label{kappa_rephrase}
\end{align}
The above structure of thermal conductivity is explicitly shown in Appendix C.

\section{Conclusion and Outlook}
We have studied diffusive spin and heat transport in a two-species atomic
Fermi gas with short-range interaction at general polarization, temperature and scattering length.
Using ``power-counting'', we first estimated the spin and thermal diffusivities in all regimes.
We suggested a method to measure the thermal diffusivity, which has not yet been experimentally measured.

Our main focus was on the magnetocaloric effects, namely the spin Seebeck and spin Peltier effects.
Observing the connection between the interaction strength and the dependence of the scattering rate on the relative momentum of two atoms,
we were able to develop a qualitative argument for the signs of magnetocaloric effects.
Near unitarity, magnetocaloric currents are in the same direction as the ``primary'' spin and heat currents,
while their directions are reversed as we move to far away from unitarity.
We then quantitatively estimated diffusivities and magnetocaloric effects in the classical regime
using approximate solutions of the Boltzmann kinetic equation,
thereby confirming the ``power-counting'' estimates of diffusivities and the sign argument for the magnetocaloric effects.
We also proved the scaling of the transport coefficients is robust to all orders of approximation, in Appendix A.

Remaining in the classical regime, we proposed an experimental procedure to detect the spin Seebeck effect
as a qualitative effect: a nonmonotonic relaxation towards equilibrium.
This method is nice in that it does not require knowledge of the equation of state of the gas.
In order to confirm these results and obtain a better understanding of the origin of these magnetocaloric effects,
we also performed a controlled perturbation to the exactly solvable Maxwellian model.
This approach agrees well with the approximate solutions of the Boltzmann equation.

To obtain the corresponding transport behavior quantitatively in the quantum degenerate (but normal) regime,
we need spin-polarized Fermi liquid theory.  We hope to report on this soon
in a forthcoming paper.

\section{Acknowledgement}

We thank Randy Hulet, Junehyuk Jung, Charles Mathy, Marco Schiro, Ariel Sommer, Henk Stoof,
and Martin Zwierlein for helpful discussions.
This work is supported by ARO Award
W911NF-07-1-0464 with funds from the DARPA OLE Program.
H. K. is partially supported by Samsung Scholarship and NSF (grant number DMR-0844115).

\appendix
\section{Approximate solution of Boltzmann equation to all orders}

\subsection{$\kappa'$ and $S_s'$}

First let's consider $\kappa'$ and $S_s'$ when $\nabla p = 0$.

Observe that $\nabla p = 0$ and mechanical equilibrium condition imply
\begin{align}
\frac{\nabla n_\uparrow}{n_\uparrow} = \frac{\nabla n_\downarrow}{n_\downarrow} = \frac{\nabla n}{n} = -\frac{\nabla T}{T} ~.
\end{align}
Therefore, Eq. (\ref{ansatz}) simplifies to
\begin{align}
\psi_{i} = \lambda\sum_{\ell=0}^{L} C_{\ell i} q^{2\ell}_i \frac{{\bf q}_i \cdot \nabla T}{T} ~,
\end{align}
where we abbreviated $C_{\ell ii} + C_{\ell ij} \equiv C_{\ell i}$.
Following the procedure explained in the main text, we can obtain the following set of linearly independent equations ($m = 1,2,\cdots,L$).
\begin{widetext}
\begin{align}
n_\uparrow \lambda^3 \left[\beta_0 C_{0\uparrow} + \beta_1 C_{1\uparrow} + \cdots \beta_L C_{L\uparrow}\right] + n_\downarrow \lambda^3 \left[\beta_0 C_{0\downarrow} + \beta_1 C_{1\downarrow} + \cdots \beta_L C_{L\downarrow}\right] & =0 \label{zero_mass}\\
\alpha_{00}\left(C_{0\uparrow} - C_{0\downarrow}\right) + \alpha_{01}\left(C_{1\uparrow} - C_{1\downarrow}\right) + \cdots +\alpha_{0L}\left(C_{L\uparrow} - C_{L\downarrow}\right) &=A_0\\
n_\downarrow \lambda^3 \left[\alpha_{m0}\left(C_{0\uparrow} - C_{0\downarrow}\right) + \left(\alpha_{m1\uparrow}C_{1\uparrow} + \alpha_{m1\downarrow}C_{1\downarrow}\right) +\cdots + \left(\alpha_{mL\uparrow}C_{L\uparrow} + \alpha_{mL\downarrow}C_{L\downarrow}\right)\right] &= A_m \\
n_\uparrow \lambda^3 \left[-\alpha_{m0}\left(C_{0\uparrow} - C_{0\downarrow}\right) + \left(\alpha_{m1\downarrow}C_{1\uparrow} + \alpha_{m1\uparrow}C_{1\downarrow}\right) +\cdots + \left(\alpha_{mL\downarrow}C_{L\uparrow} + \alpha_{mL\uparrow}C_{L\downarrow}\right)\right] &= A_m ~.
\end{align}
\end{widetext}
Here, $\{\alpha_{00},\alpha_{0 n },\alpha_{m0}, \alpha_{mn \sigma}\}$ ($m,n = 1,2,\ldots,L$ and $\sigma = \uparrow,\downarrow$)
are dimensionless numbers which may depend on $(\lambda/a)^2$ through the exponential integral and incomplete Gamma functions
and implicitly depend on the order of approximation $L$ but do not explicitly depend on temperature and densities.
$\{A_\ell, \beta_{\ell}\}$ are determined by Gaussian integrals as following:
\begin{align}
A_\ell &= \int \frac{d^3{\bf q}}{(2\pi)^3}\left(\frac{q^2}{4\pi} - \frac{5}{2}\right)q^{2\ell}q^2_z e^{-q^2/4\pi}\nonumber\\
 &= \frac{1}{3}2^{3+2\ell}\pi^{\frac{1}{2}+\ell}\ell \Gamma\left[\frac{5}{2}+\ell\right]\\
\beta_\ell &= \int \frac{d^3{\bf q}}{(2\pi)^3}q^{2\ell}q^2_z e^{-q^2/4\pi} = \frac{1}{3}2^{3+2\ell}\pi^{\frac{1}{2}+\ell} \Gamma\left[\frac{5}{2}+\ell\right] ~,
\end{align}
where $\Gamma[x]$ is the Gamma function and $A_\ell$ and $\beta_\ell$ are purely numerical numbers
independent of physical parameters.
Note that the first equation, Eq. (\ref{zero_mass}) fixes the frame to be ${\bf j}_{\uparrow} + {\bf j}_{\downarrow} = {\bf 0}$.
From dimensional analysis and the symmetry under $\uparrow \leftrightarrow \downarrow$,
we may write an ansatz solution of the above equations in the following form ($m = 1,2,\ldots,L$):
\begin{align}
C_{0\uparrow} &= -\frac{1}{\lambda^3}\left(\frac{a_0}{n_\uparrow} + \frac{b_0}{n_\downarrow} + \frac{c_0}{n}\right) \\
C_{0\downarrow} &= -\frac{1}{\lambda^3}\left(\frac{b_0}{n_\uparrow} + \frac{a_0}{n_\downarrow} + \frac{c_0}{n}\right) \\
C_{m\uparrow} &= -\frac{1}{\lambda^3}\left(\frac{a_m}{n_\uparrow} + \frac{b_m}{n_\downarrow}\right) \\
C_{m\downarrow} &= -\frac{1}{\lambda^3}\left(\frac{b_m}{n_\uparrow} + \frac{a_m}{n_\downarrow}\right) ~,
\end{align}
where $\{a_\ell, b_\ell, c_0 \}$ ($\ell = 0,1,2,\ldots,L$) are $2L+3$ unknown dimensionless numbers which may depend only on $(\lambda/a)^2$.
Since the above set of equations should hold for any values of $n_\uparrow$ and $n_\downarrow$, once we insert the ansatz to the above equations,
we obtain $2L+3$ linear equations in terms of $\{a_\ell, b_\ell, c_0\}$.
Since we are still using the same set of coefficients $\{\alpha_{00}, \alpha_{0n},\alpha_{m0},\alpha_{mn\sigma}\}$ and $\{\beta_{\ell}\}$,
it is easy to see that once the original set of equations is linearly independent
(which is a necessary condition to have an approximate solution of the Boltzmann equation),
the set of descendent equations is also linearly independent.
Therefore, once we determine all $\{a_\ell, b_\ell, c_0\}$,
the above ansatz is the ($L$'th order) approximate solution of the Boltzmann equation.

When $\nabla p =0$, we know that ${\bf j}_{spin} = -S_s' \nabla T$.
It is straightforward to show that
\begin{align}
S_s' &= -\frac{\hbar}{m}\frac{n_\uparrow}{T}\left(\sum_{\ell=0}^{L}\beta_\ell C_{\ell\uparrow}\right) \\
&= \frac{\hbar}{m}\frac{1}{T}\frac{1}{\lambda^3}\left(\sum_{\ell=0}^{L}\beta_\ell a_\ell + \frac{n_\uparrow}{n_\downarrow}\sum_{\ell=0}^{L}\beta_\ell b_\ell +\frac{n_\uparrow}{n} \beta_0 c_0\right) ~.\label{seebeck_preform}
\end{align}
Since $\{a_{\ell},b_\ell, c_0 \}$ satisfies Eq. (\ref{zero_mass}), we have two identities:
\begin{align}
\sum_{\ell=0}^{L}\beta_\ell a_\ell &= -\frac{1}{2}\beta_0 c_0 \\
\sum_{\ell=0}^{L}\beta_\ell b_\ell &= 0 ~.
\end{align}
Inserting these identities to Eq. (\ref{seebeck_preform}),
we finally obtain the full polarization and temperature dependence of the Seebeck coefficient ($\beta_0 = 2\pi$).
\begin{align}
S_s' = \frac{\hbar}{m}\frac{n_\uparrow - n_\downarrow}{n \lambda^3}\frac{\pi}{T}c_0 ~.
\end{align}
This proves that the scaling of Eq. (\ref{scaled_seebeck}) is indeed true at all orders of approximation.
Furthermore, we see that the dimensionless scaled function $h_L(x)$ ($L$ is the order of approximation) is
\begin{align}
h_L ((\lambda/a)^2) = \pi c_0 ~.
\end{align}
It should be noted that $c_0$ implicitly depends on the order of approximation.

For the thermal conductivity $\kappa'$, we are only interested in the first term, $\kappa'_1$,
which directly comes from the energy current ${\bf j}_\epsilon$.
\begin{align}
\kappa'_1 = -\frac{1}{4\pi}\frac{\hbar}{m}\sum_{\ell=0}^{L}B_\ell (n_\uparrow C_{\ell\uparrow} + n_\downarrow C_{\ell\downarrow}) ~,
\end{align}
where $B_\ell = 4\pi (5/2 + \ell)\beta_\ell$.
Plugging in ansatz solution and using Eq. (\ref{zero_mass}), we obtain $\kappa'_1$.
\begin{align}
\kappa'_1 = \frac{\hbar}{m}\frac{1}{\lambda^3}\sum_{\ell=1}^{L} \left(\frac{n_\uparrow^2 b_\ell  + 2n_\uparrow n_\downarrow a_\ell + n_\downarrow^2 b_\ell}{n_\uparrow n_\downarrow}\right)\ell \beta_\ell
\end{align}
Again, we emphasize that $a_\ell$ and $b_\ell$ are function of $(\lambda/a)^2$ and implicitly depend on the order of approximation.
This proves that the algebraic form of the polarization dependence of the thermal conductivity obtained in the main text
remains to all orders of approximation.

\subsection{$D_s$ and $P_s'$}
Now we study $D_s$ and $P_s'$ when $\nabla T = 0$ (thus $\nabla n = 0$).
As we did in the previous subsection, we want to express $\nabla n_\uparrow$ and $\nabla n_\downarrow$
in terms of $n\nabla p$.
\begin{align}
\frac{\nabla n_i}{n_i} &= \epsilon^i \frac{n\nabla p}{2n_i}~,
\end{align}
where $\epsilon^\uparrow = 1$ and $\epsilon^\downarrow = -1$.
Then, the ansatz, Eq. (\ref{ansatz}), takes the following form ($i\neq j$):
\begin{align}
\psi_i = -\epsilon^{i}\lambda\frac{n}{2}\sum_{\ell=0}^L \left(\frac{C_{\ell ii}}{n_i} - \frac{C_{\ell i j}}{n_j}\right)q_i^{2\ell}{\bf q}_i\cdot\nabla p
\end{align}
Once we define
\begin{align}
\tilde{C}_{\ell i} \equiv -\frac{n}{2}\left(\frac{C_{\ell ii}}{n_i} - \frac{C_{\ell i j}}{n_j}\right)~,
\end{align}
we reduce the system similar to the previous case.
The $2(L+1)$ linearly independent equations are ($m = 1,2,\cdots L$)
\begin{widetext}
\begin{align}
n_\uparrow \lambda^3 \left[\beta_0 {\tilde C}_{0\uparrow} + \beta_1 {\tilde C}_{1\uparrow} + \ldots \beta_L {\tilde C}_{L\uparrow}\right] - n_\downarrow \lambda^3 \left[\beta_0 {\tilde C}_{0\downarrow} + \beta_1 {\tilde C}_{1\downarrow} + \ldots \beta_L {\tilde C}_{L\downarrow}\right] & =0 \\
\alpha_{00}\left(\tilde{C}_{0\uparrow} + \tilde{C}_{0\downarrow}\right) + \alpha_{01}\left(\tilde{C}_{1\uparrow} + \tilde{C}_{1\downarrow}\right) + \cdots +\alpha_{0L}\left(\tilde{C}_{L\uparrow} + {\tilde C}_{L\downarrow}\right) &={\tilde A}_0\\
\frac{n_\uparrow n_\downarrow \lambda^3}{n} \left[\alpha_{m0}\left({\tilde C}_{0\uparrow} + {\tilde C}_{0\downarrow}\right) + \left(\alpha_{m1\uparrow}{\tilde C}_{1\uparrow} - \alpha_{m1\downarrow}{\tilde C}_{1\downarrow}\right) +\cdots + \left(\alpha_{mL\uparrow}{\tilde C}_{L\uparrow} - \alpha_{mL\downarrow}{\tilde C}_{L\downarrow}\right)\right] &= {\tilde A}_m \label{C_m_first}\\
\frac{n_\uparrow n_\downarrow \lambda^3}{n} \left[-\alpha_{m0}\left({\tilde C}_{0\uparrow} + {\tilde C}_{0\downarrow}\right) + \left(\alpha_{m1\downarrow}{\tilde C}_{1\uparrow} - \alpha_{m1\uparrow}{\tilde C}_{1\downarrow}\right) +\cdots + \left(\alpha_{mL\downarrow}{\tilde C}_{L\uparrow} - \alpha_{mL\uparrow}{\tilde C}_{L\downarrow}\right)\right] &= -{\tilde A}_m \label{C_m_second} ~.
\end{align}
\end{widetext}
$\alpha's$ and $\beta_\ell$ are same as in the previous subsection and ${\tilde A}_\ell = \beta_\ell$.

Although coefficients of Eqs. (\ref{C_m_first}) and (\ref{C_m_second}) are linearly dependent,
once we combine them, we obtain another $L$ linearly independent equations.
\begin{align}\label{constraint}
\sum_{n=1}^{L}(\alpha_{mn\uparrow} - \alpha_{mn\downarrow})({\tilde C}_{n\uparrow} - {\tilde C}_{n\downarrow})= 0 ~.
\end{align}
Together with Eq. (\ref{constraint}), we have $2(L+1)$ linearly independent equations
that uniquely determine all ${\tilde C}_{\ell i}$.

First observe that the solution of Eq. (\ref{constraint}) is trivial; ${\tilde C}_{m\uparrow} = {\tilde C}_{m\downarrow}$.
Then, we use the following ansatz \cite{ansatz}:
\begin{align}
{\tilde C}_{0\uparrow} &= -\frac{1}{\lambda^3}\left(\frac{{\tilde a}_0}{n_\uparrow} + \frac{{\tilde b}_0}{n_\downarrow}\right)\\
{\tilde C}_{0\uparrow} &= -\frac{1}{\lambda^3}\left(\frac{{\tilde b}_0}{n_\uparrow} + \frac{{\tilde a}_0}{n_\downarrow}\right)\\
{\tilde C}_{m\uparrow} &= -\frac{1}{\lambda^3}\frac{n{\tilde c}_m}{n_\uparrow n_\downarrow} = {\tilde C}_{m\downarrow} ~.
\end{align}
$\{ {\tilde a}_0, {\tilde b}_0, {\tilde c}_m\}$ are dimensionless numbers that may only depend on $(\lambda/a)^2$.
Substituting the above into original linear equations, we obtain the same linearly independent equations
in terms of $\{ {\tilde a}_0, {\tilde b}_0, {\tilde c}_m\}$.
Therefore, once we determine them, we have the approximate solution of the Boltzmann equation.

Following the same procedure when we obtained $S_s'$ and $\kappa'_1$, we can show that
\begin{align}
D_s &= \frac{\hbar}{m}\frac{2\pi}{\lambda^3}({\tilde a}_0 - {\tilde b}_0) \\
P_{s1}' &= \frac{\hbar}{m}\frac{T}{n\lambda^3}\sum_{\ell=1}^{L}\left(\frac{n_\uparrow^2 - n_\downarrow^2}{n_\uparrow n_\downarrow}\right){\tilde c}_\ell \ell \beta_\ell ~,
\end{align}
where $P_{s1}'$ is the first term in the Peltier coefficient.
Once we scale the scattering length with $\lambda$,
the polarization and temperature dependence of $D_s$ and $P_{s1}'$ at an arbitrary $L$
remains the same as for $L=1$.
This proves that the scaling structure of the transport coefficients remains the same to all orders of approximation.

\section{Calculation of the spin Seebeck effect}
Here we present derivations of signatures of the spin Seebeck effect in detail.
First, apply a long wavelength temperature modulation to the system.
The mechanical equilibrium condition implies that a temperature modulation is equivalent to a total density modulation.
Initially the gas is uniformly polarized so that the density of each spin component is
\begin{align}
n_{i}(t=0,z) &= n_{i 0} + \delta n_{i 0}\cos wz\\
\delta n_{i 0} &= \frac{n_{i 0}}{n_{0}}\delta n_{0}~\label{species_density_deviation},
\end{align}
where $\delta n_0$ is the maximum value of total density deviation from the average density.
In order to calculate the change of densities of each species,
which is directly measurable in experiment and contains the signature of the spin Seebeck effect,
we need the space-time dependence of the density of each species.
We will use the diffusion matrix and the continuity equation to derive the space-time dependence of densities.
Since the continuity equation for the particle number density is $\partial_t n_i + \nabla \cdot {\bf j}_i = 0$,
it is practical to write diffusion equation in the format of Eq. (\ref{diffusion}).
With this initial condition, a nonuniform particle current flows, but in this classical limit, there is no net energy current.
Thus we work in the reference frame where the local energy current vanishes
(${\bf j}_{\epsilon \uparrow} + {\bf j_{\epsilon \downarrow}} = 0$).
Thanks to Galilean invariance, this
choice of a reference frame does not affect the physics.

Following a similar procedure as described in the main text, but in this zero energy current frame,
we can determine all coefficients $C_{\ell ij}$ ($\ell = 0,1$)
and express the particle current of each species in terms of $\nabla n_\uparrow$ and $\nabla n_\downarrow$.
Then, we can write currents in the diffusion matrix format as in Eq. (\ref{diffusion}):
\begin{widetext}
Near unitarity,
\begin{equation}\label{diffusion_unitarity}
\begin{pmatrix}
\mathbf{j}_{\uparrow}\\
\mathbf{j}_{\downarrow}
\end{pmatrix}
=-\frac{9\pi^{3/2}}{9728\sqrt{2}}\left(\frac{\hbar}{m}\left(\frac{T}{T_{F}}\right)^{3/2}\right)
\begin{pmatrix}
\frac{365 n_{\uparrow}^{2} + 354n_{\uparrow}n_{\downarrow} + 733n_{\downarrow}^{2}}{n n_{\downarrow}} &&
\frac{381 n_{\uparrow}^{2} + 42n_{\uparrow}n_{\downarrow} + 405n_{\downarrow}^{2}}{n n_{\downarrow}}\\
\frac{405 n_{\uparrow}^{2} + 42n_{\uparrow}n_{\downarrow} + 381n_{\downarrow}^{2}}{n n_{\uparrow}} &&
\frac{733 n_{\uparrow}^{2} + 354n_{\uparrow}n_{\downarrow} + 365n_{\downarrow}^{2}}{n n_{\uparrow}}
\end{pmatrix}
\begin{pmatrix}
\nabla n_{\uparrow}\\
\nabla n_{\downarrow}
\end{pmatrix}~.
\end{equation}
Far from unitarity,
\begin{equation}\label{diffusion_away_from_unitarity}
\begin{pmatrix}
\mathbf{j}_{\uparrow}\\
\mathbf{j}_{\downarrow}
\end{pmatrix}
=-\frac{3\pi^{3/2}}{896\sqrt{2}}\left(\frac{\hbar}{m (k_F a)^2}\sqrt{\frac{T}{T_{F}}}\right)
\begin{pmatrix}
\frac{145 n_{\uparrow}^{2} + 158n_{\uparrow}n_{\downarrow} + 289n_{\downarrow}^{2}}{n n_{\downarrow}} &&
\frac{137 n_{\uparrow}^{2} - 14n_{\uparrow}n_{\downarrow} + 125n_{\downarrow}^{2}}{n n_{\downarrow}}\\
\frac{125 n_{\uparrow}^{2} - 14n_{\uparrow}n_{\downarrow} + 137n_{\downarrow}^{2}}{n n_{\uparrow}} &&
\frac{289 n_{\uparrow}^{2} + 158n_{\uparrow}n_{\downarrow} + 145n_{\downarrow}^{2}}{n n_{\uparrow}}
\end{pmatrix}
\begin{pmatrix}
\nabla n_{\uparrow}\\
\nabla n_{\downarrow}
\end{pmatrix}~.
\end{equation}
These matrix equations contain two eigenmodes,
one is the thermal mode with eigenvalue ${\frak D}_{T}$ ($\mathbf{j}_{\uparrow}$ and $\mathbf{j}_{\downarrow}$ are in the same directions)
and the other is the spin mode with eigenvalue ${\frak D}_{s}$ ($\mathbf{j}_{\uparrow}$ and $\mathbf{j}_{\downarrow}$ are in opposite directions).
As expected, ${\frak D}_T > {\frak D}_s$ for all polarizations in both of these limits.
Applying continuity equations to the above diffusion matrix equations
while keeping all differential operators linear
(we restrict ourselves in a linear response theory), we obtain two-component diffusion equations:
\begin{equation}
\frac{\partial}{\partial t}
\begin{pmatrix}
n_{\uparrow}\\
n_{\downarrow}
\end{pmatrix}
=-
\begin{pmatrix}
{\frak D}_{\uparrow\uparrow} && {\frak D}_{\uparrow\downarrow}\\
{\frak D}_{\downarrow\uparrow} && {\frak D}_{\downarrow\downarrow}
\end{pmatrix}
\begin{pmatrix}
\nabla^2 n_{\uparrow}\\
\nabla^2 n_{\downarrow}
\end{pmatrix}~.
\end{equation}
Therefore, we can immediately write the time evolution of each species in terms of the eigenmodes as
\begin{equation}\label{time_evolution}
\begin{pmatrix}
n_{\uparrow}(t,z)\\
n_{\downarrow}(t,z)
\end{pmatrix}
=
\begin{pmatrix}
n_{\uparrow 0}\\
n_{\downarrow 0}
\end{pmatrix}
+\alpha e^{-{\frak D}_{s}w^{2}t}\delta n_{\downarrow 0}
\begin{pmatrix}
\gamma \\
1
\end{pmatrix}
\cos wz
+
\beta e^{-{\frak D}_{T}w^{2}t}\delta n_{\downarrow 0}
\begin{pmatrix}
\zeta \\
1
\end{pmatrix}
\cos wz.
\end{equation}
\end{widetext}
($\gamma$, 1) and ($\zeta$, 1) are the eigenvectors of the spin mode and the thermal mode, respectively.
$\gamma$ is negative while $\zeta$ is positive.
$\alpha$ and $\beta$ are determined by the initial condition,
\begin{equation}
\begin{pmatrix}
\delta n_{\uparrow 0}\\
\delta n_{\downarrow 0}
\end{pmatrix}
=
\alpha \delta n_{\downarrow 0}
\begin{pmatrix}
\gamma\\\
1
\end{pmatrix}
+
\beta \delta n_{\downarrow 0}
\begin{pmatrix}
\zeta\\
1
\end{pmatrix}.
\end{equation}
It is straightforward to derive $\gamma$, $\zeta$, $\alpha$, $\beta$, ${\frak D}_{s}$ and ${\frak D}_T$.
For example,
\begin{align}
\alpha &= \frac{1}{\gamma - \zeta}\left(\frac{n_{0\uparrow}}{n_{0\downarrow}} - \zeta\right)\\
\beta &= 1- \alpha ~.
\end{align}
Explicit expressions for $\gamma$, $\zeta$, ${\frak D}_s$
and ${\frak D}_T$ are fairly lengthy so they are omitted here.

The spin Seebeck effect is most apparent when observing the densities at the edges of the system ($wz = 0$ or $wz = \pi$)
so we choose to focus on the cold edge, $z = 0$.
The uniform pressure condition implies that both majority and minority densities are initially higher than average at the cold side.
Thus, initially $\delta n_{\downarrow} (t=0, z=0) > 0$.
Even though the initial condition can be chosen to be the same for both unitarity and far away from unitarity,
the time evolution of density of each species is qualitatively different for these two limiting cases.
As argued in the main text, the minority density deviation has an extremum before it relaxes
to zero at unitarity whereas the majority density deviation has an extremum far away from unitarity.
Therefore, a change in the sign of a certain species is a unique signature of
the spin Seebeck effect.
We can express the magnitude of the spin Seebeck signal in dimensionless form by
$\frac{\delta n_i(t,0)}{\delta n_{i0}}$.
This quantity is plotted in figures \ref{deviation_unitarity} and \ref{deviation_away}.
At time $t_{i,ext}$ ($i = \downarrow$ near unitarity and $i = \uparrow$ away from unitarity), the density of the species $i$ reaches an extremum.
It is easy to derive formal expressions of $t_{i,ext}$ and a dimensionless measure, $\left|\frac{\delta n_i(t_{i,ext},0)}{\delta n_{i0}}\right|$.

Near unitarity,
\begin{align}
t_{\downarrow,ext} &= \frac{1}{({\frak D}_T - {\frak D}_s)w^{2}}\ln\left|\frac{{\frak D}_T\beta}{{\frak D}_{s}\alpha}\right|\\
\left|\frac{\delta n_{\downarrow}(t_{\downarrow,ext},0)}{\delta n_{\downarrow0}}\right| &= \alpha \exp\left[-\frac{{\frak D}_s}{{\frak D}_T - {\frak D}_s}\ln\left|\frac{{\frak D}_T\beta}{{\frak D}_s\alpha}\right|\right]\nonumber\\
&+ \beta \exp\left[-\frac{{\frak D}_T}{{\frak D}_T - {\frak D}_s}\ln\left|\frac{{\frak D}_T\beta}{{\frak D}_s\alpha}\right|\right] ~.
\end{align}
Away from unitarity,
\begin{align}
t_{\uparrow,ext} &= \frac{1}{({\frak D}_T - {\frak D}_s)w^{2}}\ln\left|\frac{{\frak D}_T\beta\zeta}{{\frak D}_{s}\alpha\gamma}\right|\\
\left|\frac{\delta n_{\uparrow}(t_{\uparrow,ext},0)}{\delta n_{\uparrow0}}\right| &= \frac{\alpha\gamma}{\alpha\gamma + \beta\zeta} \exp\left[-\frac{{\frak D}_s}{{\frak D}_T - {\frak D}_s}\ln\left|\frac{{\frak D}_T\beta\zeta}{{\frak D}_s\alpha\gamma}\right|\right] \nonumber\\
&+ \frac{\beta\zeta}{\alpha\gamma + \beta\zeta} \exp\left[-\frac{{\frak D}_T}{{\frak D}_T - {\frak D}_s}\ln\left|\frac{{\frak D}_T\beta\zeta}{{\frak D}_s\alpha\gamma}\right|\right]~.\nonumber\\
\end{align}

$\left|\frac{\delta n_i(t_{i,ext},0)}{\delta n_{i0}}\right|$ is plotted in figure \ref{delta_n_ext}.
As expected, the unitarity limit shows a stronger signal of the spin Seebeck effect than far away from unitarity.

Another consequence of the spin Seebeck effect is the change of polarization.
From Eqs. (\ref{species_density_deviation}) and (\ref{time_evolution}), we can express the polarization as a function of time.
Keeping only linear term of $\delta n_0$, we obtain
\begin{align}
\delta p(t,0) &= p(t,0) - p_0 \nonumber\\
&=\frac{\delta n_0}{n_0}\bigg[\frac{n_{\downarrow 0}}{n_0}\left(\alpha (\gamma -1)e^{-{\frak D}_s w^2 t}+\beta(\zeta -1)e^{-{\frak D}_T w^2 t}\right)\nonumber\\
&\quad - p_0 \frac{n_{\downarrow 0}}{n_0}\left(\alpha (\gamma + 1)e^{-{\frak D}_s w^2 t}+\beta(\zeta + 1)e^{-{\frak D}_T w^2 t}\right)\bigg] ~,
\end{align}
where $p_0 = (n_{\uparrow 0} - n_{\downarrow 0})/n_0$.
Figure \ref{polarization} is the plot of $\delta p(t,0)$ normalized by $\delta n_0/n_0$.

\section{Manifestation of the Onsager relation}
For completeness, we present diffusion matrices of approximate solutions
and the solution of the first order perturbation of the Maxwellian model
in the form of Eq. (\ref{diffusion_Tmu}) where the Onsager relation should be explicit.
We simply transform the set of driving forces $(\nabla T, n\nabla p)$
to another set of driving forces $(\nabla T, \nabla (\mu_\uparrow - \mu_\downarrow))$
associated with the entropy production.

\begin{widetext}
For approximate solutions we have the following:

At unitarity ($|a|\gg \lambda$),
\begin{equation}\label{unitarity_zeromass_onsager}
\begin{pmatrix}
\mathbf{j}_{heat}\\
\mathbf{j}_{spin}
\end{pmatrix}
=-\frac{45\pi^{3/2}}{608\sqrt{2}}\left(\frac{\hbar}{m}\left(\frac{T}{T_{F}}\right)^{3/2}\right)
\begin{pmatrix}
\kappa_u && (n_\uparrow - n_\downarrow) - \frac{39 n_\uparrow n_\downarrow}{5n} \ln \left(\frac{n_\uparrow}{n_\downarrow}\right) \\
\frac{(n_\uparrow - n_\downarrow)}{T} - \frac{39 n_\uparrow n_\downarrow}{5n T} \ln \left(\frac{n_\uparrow}{n_\downarrow}\right) && \frac{39 n_\uparrow n_\downarrow}{5n T}
\end{pmatrix}
\begin{pmatrix}
\nabla T\\
\nabla (\mu_\uparrow - \mu_\downarrow)
\end{pmatrix}~,
\end{equation}
where $\kappa_u$ is
\begin{equation}
\kappa_u = \frac{5}{16}n\left(\frac{(73n_{\uparrow}^{2} + 82n_{\uparrow}n_{\downarrow} + 73n_{\downarrow}^{2})}{n_{\uparrow}n_{\downarrow}}\right) - \ln\left(\frac{n_\uparrow}{n_\downarrow}\right)\left(\frac{10(n_\uparrow^2 - n_\downarrow^2) - 39 n_\uparrow n_\downarrow \ln (n_\uparrow/n_\downarrow)}{5n}\right).
\end{equation}
Far away from unitarity ($|a| \ll \lambda$),
\begin{equation}\label{away_zeromass_onsager}
\begin{pmatrix}
\mathbf{j}_{heat}\\
\mathbf{j}_{spin}
\end{pmatrix}
=-\frac{15\pi^{3/2}}{224\sqrt{2}}\left(\frac{\hbar}{m (k_{F}a)^{2}}\sqrt{\frac{T}{T_{F}}}\right)
\begin{pmatrix}
\kappa_a && -(n_\uparrow - n_\downarrow) - \frac{43 n_\uparrow n_\downarrow}{5n} \ln \left(\frac{n_\uparrow}{n_\downarrow}\right) \\
-\frac{(n_\uparrow - n_\downarrow)}{T} - \frac{43 n_\uparrow n_\downarrow}{5nT} \ln \left(\frac{n_\uparrow}{n_\downarrow}\right) && \frac{43 n_\uparrow n_\downarrow}{5n T}
\end{pmatrix}
\begin{pmatrix}
\nabla T\\
\nabla (\mu_\uparrow - \mu_\downarrow)
\end{pmatrix}~,
\end{equation}
where $\kappa_a$ is
\begin{equation}
\kappa_a = \frac{5}{8}n\left(\frac{(29n_{\uparrow}^{2} + 26n_{\uparrow}n_{\downarrow} + 29n_{\downarrow}^{2})}{n_{\uparrow}n_{\downarrow}}\right) + \ln\left(\frac{n_\uparrow}{n_\downarrow}\right)\left(\frac{10(n_\uparrow^2 - n_\downarrow^2) + 43n_\uparrow n_\downarrow \ln (n_\uparrow/n_\downarrow)}{5n}\right).
\end{equation}

For the first order perturbation of the Maxwellian model, we have the following:
\begin{align}\label{maxwell_zeromass_onsager}
\begin{pmatrix}
\mathbf{j}_{heat}\\
\mathbf{j}_{spin}
\end{pmatrix}
=-D_{0}
\begin{pmatrix}
\kappa_M
&& -\frac{\pi\epsilon (n_{\uparrow} - n_{\downarrow})}{ \lambda^{2} } - \frac{n_\uparrow n_\downarrow}{5n }\left(1 - \frac{20\pi\epsilon}{\lambda^{2}}\right)\ln\left(\frac{n_\uparrow}{n_\downarrow}\right)\\
-\frac{\pi\epsilon (n_{\uparrow} - n_{\downarrow})}{\lambda^{2} T} - \frac{n_\uparrow n_\downarrow}{5n T}\left(1 - \frac{20\pi\epsilon}{\lambda^{2}}\right)\ln\left(\frac{n_\uparrow}{n_\downarrow}\right) && \frac{n_\uparrow n_\downarrow}{5n T}\left(1 - \frac{20\pi\epsilon}{\lambda^{2}}\right)
\end{pmatrix}
\begin{pmatrix}
\nabla T \\
\nabla (\mu_\uparrow - \mu_\downarrow)
\end{pmatrix}~,
\end{align}
where $\kappa_M$ is
\begin{equation}
\kappa_M = \frac{1}{2}n\left(\frac{(n_{\uparrow}^{2} + n_{\uparrow}n_{\downarrow} + n_{\downarrow}^{2})}{n_{\uparrow}n_{\downarrow}} - \frac{3\pi\epsilon}{\lambda^{2}}\frac{(9n_{\uparrow}^{2} + 10n_{\uparrow}n_{\downarrow} + 9n_{\downarrow}^{2})}{n_{\uparrow}n_{\downarrow}}\right) + \frac{2 \pi \epsilon \ln\left(\frac{n_\uparrow}{n_\downarrow}\right)}{n \lambda^2} \left((n_\uparrow^2 - n_\downarrow^2) + n_\uparrow n_\downarrow\left( \frac{\lambda^2}{10\pi\epsilon} -2\right) \ln\left(\frac{n_\uparrow}{ n_\downarrow}\right)\right)
\end{equation}
\end{widetext}

The above matrices clearly exhibit the Onsager relation.
Also each off-diagonal element carries a term proportional to $-\sigma_s \ln(n_\uparrow / n_\downarrow)$
as mentioned in Eq. (\ref{seebeck_kubo_log}) in section VI.
Lastly, the structure of thermal conductivity (Eq. (\ref{kappa_rephrase})) is evident.


\begin{thebibliography}{99}
\bibitem{energy}
In the limit of high temperature, a gas is still dilute enough so that kinetic energy gives the dominant contribution to the total energy.
This is a necessary condition to use the Boltzmann approach.
\bibitem{ansatz}
Motivation of this choice came from the observation that the right hand side of the linear equation does not depend on densities.
Therefore, a quantity in the big parenthesis in the left hand side must be proportional to $n/(n_\uparrow n_\downarrow)$.
\end{thebibliography}
\end{document}